\documentclass[a4paper,fleqn]{cas-sc}

\usepackage[authoryear]{natbib}

\def\tsc#1{\csdef{#1}{\textsc{\lowercase{#1}}\xspace}}
\tsc{WGM}
\tsc{QE}

\newcommand{\ch}{CH$_4$}
\newcommand{\chd}{CH$_3$D}
\newcommand{\ac}{C$_2$H$_2$}
\newcommand{\acd}{C$_2$HD}
\newcommand{\ethy}{C$_2$H$_4$}
\newcommand{\ethyd}{C$_2$H$_3$D}
\newcommand{\cm}{cm$^{-1}$}
\newcommand{\dg}{$^{\circ}$}

\begin{document}
\let\WriteBookmarks\relax
\def\floatpagepagefraction{1}
\def\textpagefraction{.001}

\shorttitle{D/H ratio in Titan's acetylene}    

\shortauthors{<B. B\'ezard et al.>}  

\title [mode = title]{The D/H ratio in Titan's acetylene from high spectral resolution IRTF/TEXES observations}  



%

\author[1]{B. B\'ezard}[orcid=0000-0002-5433-5661]

\cormark[1]


\ead{bruno.bezard@obspm.fr}


\affiliation[1]{organization={LESIA, Observatoire de Paris, Universit\'e PSL, Sorbonne Universit\'e, Universit\'e Paris Cit\'e, CNRS},
            addressline={Place Jules Janssen}, 
            city={Meudon},
           citysep={}, 
            postcode={92195}, 
            country={France}}

\author[2]{C.A. Nixon}
\ead{conor.a.nixon@nasa.gov}

\author[1]{S. Vinatier}
\ead{sandrine.vinatier@obspm.fr}

\author[1]{E. Lellouch}
\ead{emmanuel.lellouch@obspm.fr}

\author[3]{T. Greathouse}
\ead{thomas.greathouse@swri.org}
\fnmark[1]

\author[3]{R. Giles}
\ead{rohini.giles@swri.org}

\author[2]{N.A. Lombardo}
\ead{nicholas.lombardo@yale.edu}
\fnmark[2]

\author[4]{A. Jolly}
\ead{jolly@lisa.ipsl.fr}

\affiliation[2]{organization={NASA Goddard Space Flight Center},
            city={Greenbelt},
            postcode={20771}, 
            state={MD},
            country={United States}}
            
 \affiliation[3]{organization={Southwest Research Institute},
            city={San Antonio},
            citysep={}, 
            state={TX},
            country={United States}}
            
\affiliation[4]{organization={LISA, CNRS UMR 7583, Universit\'e Paris-Est Cr\'eteil, Universit\'e Paris-Cit\'e,  IPSL},
            city={Cr\'eteil},
            citysep={}, 
            postcode={94010}, 
            country={France}}

\cortext[1]{Corresponding author}

\fntext[1]{Visiting Astronomer at the Infrared Telescope Facility, which is operated by the University of Hawaii under contract 80HQTR19D0030 with the National Aeronautics and Space Administration}
\fntext[2]{Present address: Department of Earth and Planetary Sciences,Yale University, New Heaven, CT 06511, United States}

\nonumnote{}

\begin{abstract}
 We report observations of deuterated acetylene (\acd) at 19.3 $\mu$m (519 \cm) with the Texas Echelon Cross Echelle Spectrograph on the NASA Infrared Telescope Facility in July 2017. Six individual lines from the Q-branch of the $\nu_4$ band were clearly detected with a S/N ratio up to 10. Spectral intervals around 8.0 $\mu$m (745 \cm) and 13.4 $\mu$m (1247 \cm) containing  acetylene (\ac) and methane (\ch) lines respectively, were observed during the same run to constrain the disk-averaged  \ac\ abundance profile and temperature profile. {\em Cassini} observations with the Composite Infrared Spectrometer (CIRS) were used to improve the flux calibration and help to constrain the atmospheric model. The measured D/H ratio in acetylene, derived from the \acd/\ac\ abundance ratio, is (1.22$^{+0.27}_{-0.21})$~$\times$ 10$^{-4}$, consistent with that in methane obtained in previous studies. Possible sources of fractionation at different steps of the acetylene photochemistry are investigated.
  
\end{abstract}



\begin{keywords}
Titan  \sep Planetary atmospheres  \sep Isotopic abundances \sep Atmospheric composition  \sep Infrared spectroscopy
\end{keywords}

\maketitle

\section{Introduction}\label{Intro}

Titan's complex photochemistry produces a suite of hydrocarbons and nitrogen-bearing compounds, eventually leading to the formation of haze particles. This chemistry is initiated by the dissociation and ionization of dinitrogen (N$_2$) and methane (CH$_4$) by ultraviolet solar radiation, particles from Saturn's magnetosphere and galactic cosmic rays. The vertical and horizontal abundance profiles of many of these photochemical species along with their seasonal variations have been extensively characterized through Cassini observations \citep[e.g.][for the most recent ones]{Coustenis2020, Mathe2020, Tribett2021, Vinatier2020}. These investigations are complemented by other space-based and ground-based millimeter and mid-infrared observations, which in particular allowed the detections of new photochemical compounds \citep{Lombardo2019, Moreno2011, Nixon2020, Thelen2020}. 

Observational data serve as constraints for photochemical models that aim at understanding the chemistry at work in Titan's atmosphere \citep[e.g.][]{Moreno2012, Dobrijevic2014, Krasnopolsky2014,  Dobrijevic2016, Lara2014, Vuitton2019}. Besides chemical abundance profiles, isotopic ratios may provide valuable information on the chemical and physical processes involved in the production and loss of the species. For example, the large difference in the $^{14}$N/$^{15}$N ratio observed in HCN \citep{Marten2002, Gurwell2004, Vinatier2007b, Courtin2011} and in N$_2$ \citep{Niemann2010}, the main nitrogen reservoir, allows us to constrain the relative flux of fractionated N atoms from photolysis and non-fractionated N atoms from other processes, such as magnetospheric electrons or galactic cosmic rays \citep{Dobrijevic2018, Vuitton2019}. 

Dissociation of methane by solar ultraviolet radiation is the primary source of the hydrocarbons produced on Titan. Methane is also lost by reactions with various photochemical products and, most importantly, by reaction with the C$_2$H radical below 600 km. These processes are potentially a source of hydrogen fractionation in the photochemical products as they preferentially break the C-H bond over the stronger C-D bond \citep{Pinto1986, Lunine1999}. Thus measuring the D/H ratio in various hydrocarbons can shed light on different photochemical pathways at work on Titan.

Besides methane, the only hydrocarbon in which the D/H ratio has been measured is acetylene (\ac). Fitting simultaneously the 14.7-$\mu$m (678 \cm) $\nu_5$ band of \acd\ and the nearby 13.7-$\mu$m (729 \cm) $\nu_5$ band of \ac\ observed in spectral averages of nadir Cassini/CIRS spectra, \cite{Coustenis2008} inferred a D/H ratio in \ac\ of (2.09$\pm$0.45)~$\times$ 10$^{-4}$. This ratio appears significantly larger than that in methane, which is in the range (1.1--1.6)~$\times$ 10$^{-4}$  \citep{Owen1986, deBergh1988, Coustenis1989, Coustenis2003, Penteado2005, Bezard2007, Coustenis2007, Abbas2010, deBergh2012, Nixon2012}. However, from the analysis of CIRS limb spectra, \cite{Coustenis2008} derived a lower value of the D/H ratio in \ac, (1.63$\pm$0.27)~$\times$ 10$^{-4}$, closer to the value in \ch. It can be noted that the \acd\ emission feature is weak and mixed with \ac\ emission features at the CIRS resolution (0.5 \cm). \cite{Coustenis2008} also detected a weak emission from the $\nu_4$ band of \acd\ centered at 19.3 $\mu$m (519 \cm) in a large average of CIRS spectra (mixing disk and limb data). However, they did not analyze it due to the low S/N ratio and the presence of spurious instrumental features.

To overcome these difficulties and improve the precision in the D/H in acetylene, we used the Texas Echelon Cross Echelle Spectrograph (TEXES) on the NASA Infrared Telescope Facility (IRTF) to observe the $\nu_4$ band of \acd\  at an unprecedented resolving power $\lambda / \Delta \lambda$ of $\sim$65,000. We also observed spectral ranges around 8.0 and 13.4 $\mu$m to constrain the temperature profile and the \ac\ abundance profile in our radiative transfer analysis, in order to improve the accuracy of the \acd / \ac\ ratio determination.

\section{TEXES Observations}\label{Texes_obs}

Observations of Titan were conducted with TEXES \citep{Lacy2002} mounted at the NASA/IRTF on July 8 and 14, 2017 UT under programmes 2017A045 (PI: B\'ezard) and 2017A109 (PI: Nixon) respectively. We targeted three spectral intervals containing lines from \acd\ (518.4--520.0 \cm), \ac\ (742.9--746.7 \cm) and \ch\ (1244.3-1250.7 \cm). TEXES was used in the high-resolution cross-dispersed mode, achieving a resolving power between 65,000 and 95,000. Details of the observations are given in Table~\ref{table_obs}. Titan's angular size was 0.78 arcsec, less than the slit width (2.0 or 1.4 arcsec), so that our measurements pertain to disk-averaged conditions. The sub-Earth and sub-Sun latitudes were both 27\dg N. Observations occurred around Titan's summer solstice.

 \begin{table}
\caption{TEXES Observations of Titan}\label{table_obs}
\begin{tabular*}{\tblwidth}{@{}LLLLLLLL@{}}
\toprule
Target  &  Date & Spectral & Spectral & Titan--Earth & Slit width & Integration & Maximum S/N \\ 
molecule & (UT) & interval (\cm) & resolution (\cm) & distance (AU) & (arcsec) & time (s) & (per spectel) \\ 
\midrule
\acd & July 8, 2017 & 518.4--520.0 & 0.0077 & 9.124 & 2.0 & 910 & 15 \\
\ac & July 14, 2017 & 742.9--746.7 & 0.0078 & 9.163 & 1.4 & 830 & 50 \\
\ch & July 8, 2017 & 1244.3--1250.7 & 0.0156 & 9.124 & 1.4 & 770 & 20 \\
\bottomrule
\end{tabular*}
\end{table}

The \acd\ observations were performed by nodding Titan 5 arcsec along the 2.0~$\times$~15 arcsec slit in order to remove the sky emission. For the \ac\ and \ch\ settings, the nodding amplitude was 3 arsec along the 1.4~$\times$~9 and 1.4~$\times$~7 arcsec slits respectively. The individual spectra were flat fielded, calibrated and co-added using the TEXES pipeline software package with the procedure described in \cite{Lacy2002}. To remove the residual telluric absorption, we divided the Titan spectra by spectra of the asteroid 10~Hygiea recorded shortly after. 

In a first step, we also relied on the 10~Hygiea spectra to obtain a flux calibration of our Titan observations. To do so, we used the asteroid Standard Thermal Model (STM) \citep{Lebofsky1989} with the Earth distance, Sun distance and phase angle corresponding to our observations. We used the asteroid radius determined by \cite{Vernazza2021} (216.5 km) and the beaming parameter determined by \cite{Lim2005} (0.81). The Bond albedo and emissivity were taken as 0.07 and 0.98 respectively, as in \cite{Lim2005}. The STM spectrum, calculated for zero solar phase angle, was corrected for phase by 0.01 magnitude per degree (10~Hygiea's phase angle was 3.5\dg\ for the July 8 observations and 5.8\dg\ for the July 14 observations). Titan's flux (in Jy) is then given by:
\begin{equation}\label{fluxTexes}
F_T= \frac{S_T}{S_a}F_a  f,
\end{equation}
where $S_T$ and $S_a$ are respectively the Titan and asteroid measured fluxes (in TEXES units), $F_a$ is the asteroid flux (in Jy) and $f$ a possible corrective factor (see Section~\ref{CIRS_obs}), for now set to 1.

\begin{figure}
	\centering
		\includegraphics[trim={0 1.0cm 0 0},clip,width=0.7\columnwidth]{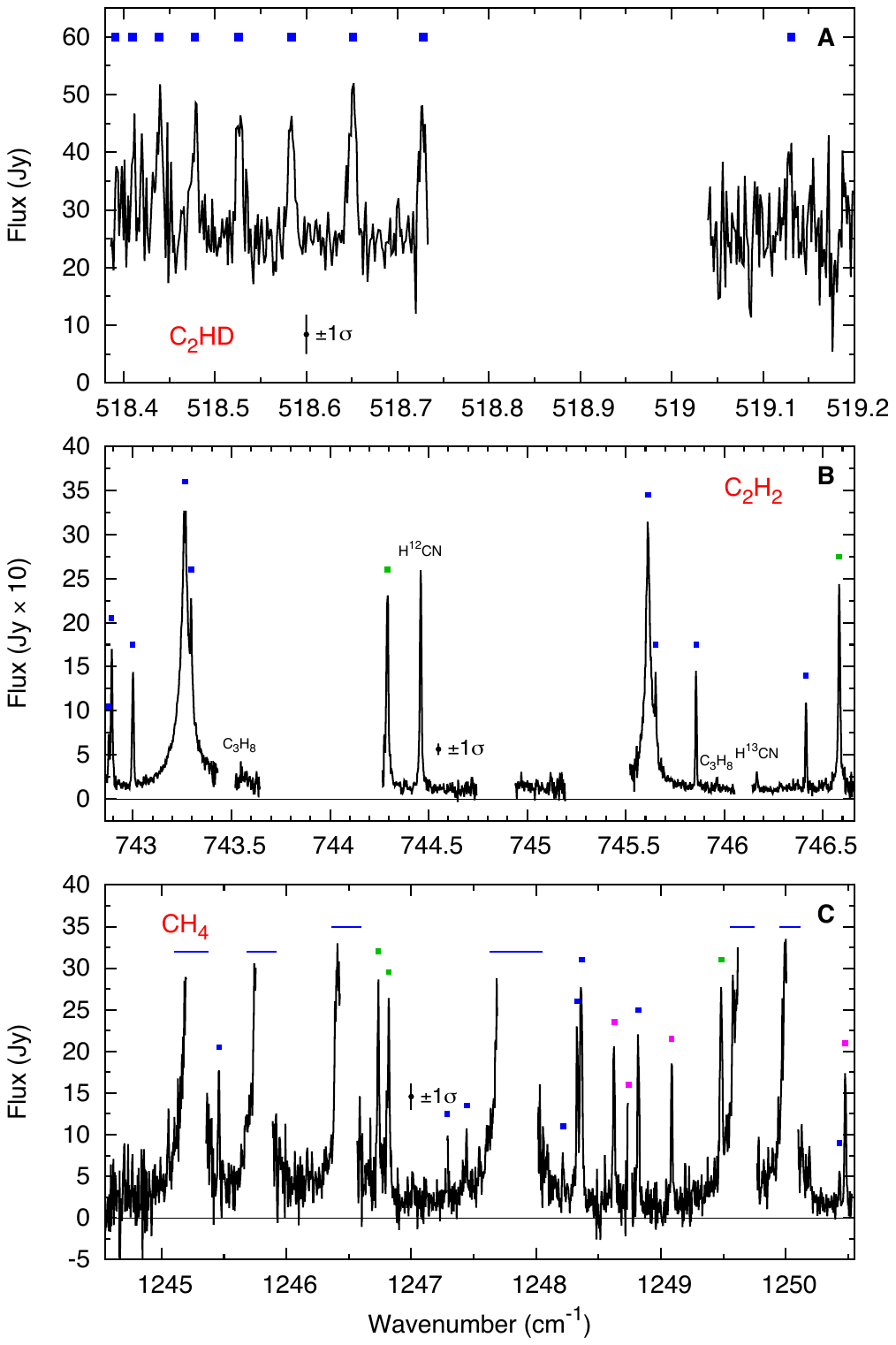}
	  \caption{TEXES spectra recorded in three spectral intervals containing lines of \acd\ (A), \ac\ (B) and \ch\ (C). Spectra have been flux-calibrated against 10~Hygiea observations using the asteroid STM (see text). In panel A, the locations of the \acd\ lines are indicated as blue squares. In panel B, all the detected lines are from \ac, except for two HCN lines and two weak C$_3$H$_8$ features which are labeled. Blue squares indicate $^{12}$C$^{12}$CH$_2$ lines and green squares  $^{12}$C$^{13}$CH$_2$ lines. In panel C, all lines are due to methane. Besides strong $^{12}$CH$_4$ multiplets (horizontal blue lines), weaker lines from $^{12}$CH$_4$ (blue squares), $^{13}$CH$_4$ (green squares) and \chd\ (purple squares) are clearly detected. The $\pm$1$\sigma$ noise error bars are indicated. Missing intervals in the data correspond to either troughs between grating orders or regions with low telluric transmission.}
	  \label{Fig_obs_Texes}
\end{figure}

Figure~\ref{Fig_obs_Texes} shows the TEXES spectra in the three settings. In the \acd\ setting, we clearly detect several lines from the Q-branch of the \acd\ $\nu_4$ band with a signal-to-noise (S/N) ratio up to 10. These lines appear on a continuum mostly due to N$_2$--\ch\ and N$_2$--H$_2$ collision-induced absorption and to haze emission. The \ac\ setting exhibits \ac\ lines of different intensities along with a H$^{12}$CN line, a weak H$^{13}$CN line and two weak emission features from C$_3$H$_8$. The \ch\ setting shows many strong and weak methane lines. The Titan-Earth Doppler shift of 15 km s$^{-1}$ allowed us to observe the full low-frequency half of the Titan \ch\ strong lines. In all plots, regions of low telluric transmission have been removed.

\section{Cassini/CIRS observations}\label{CIRS_obs}

Because the \acd, \ac\ and \ch\ lines were not recorded simultaneously in the TEXES spectra, the precision of the D/H ratio we are able to achieve is sensitive to any error in the relative flux calibration of the three settings. We expect that our flux calibration, based on the ratio between Titan and 10~Hygiea observations, is no more than 10-15 \% accurate. Error sources come from possible flux variations of asteroid 10~Hygiea with rotational phase and possible spectral variations of its emissivity between the three TEXES settings. Also, variations in the atmospheric seeing and pointing errors can induce flux calibration errors.

To get around this difficulty, we employed spectra recorded by the Cassini Composite Infrared Spectrometer \citep[CIRS;][]{Flasar2004, Jennings2017} to calibrate our TEXES data. CIRS was a Fourier transform spectrometer, composed of three focal planes covering the spectral ranges 10--600 \cm\ (FP1), 580--1100 \cm\ (FP3) and 1050--1500 \cm\ (FP4) with a spectral resolution adjustable between 0.5 and 15 \cm. FP1 had a single detector with a circular field of view (FOV) of 3.9-mrad diameter while FP3 and FP4 each included a linear array of ten detectors providing an instantaneous field of view (IFOV) of 0.27 mrad (square). The CIRS absolute radiometric calibration is expected to be precise at  the 1\% level \citep{Jennings2017} but great care must be taken in selecting the CIRS spectra for comparison with our disk-averaged TEXES observations.

\begin{figure}
	\centering
		\includegraphics[angle=-90,trim={1.5cm 0 2.5cm 0},clip,width=1.\columnwidth]{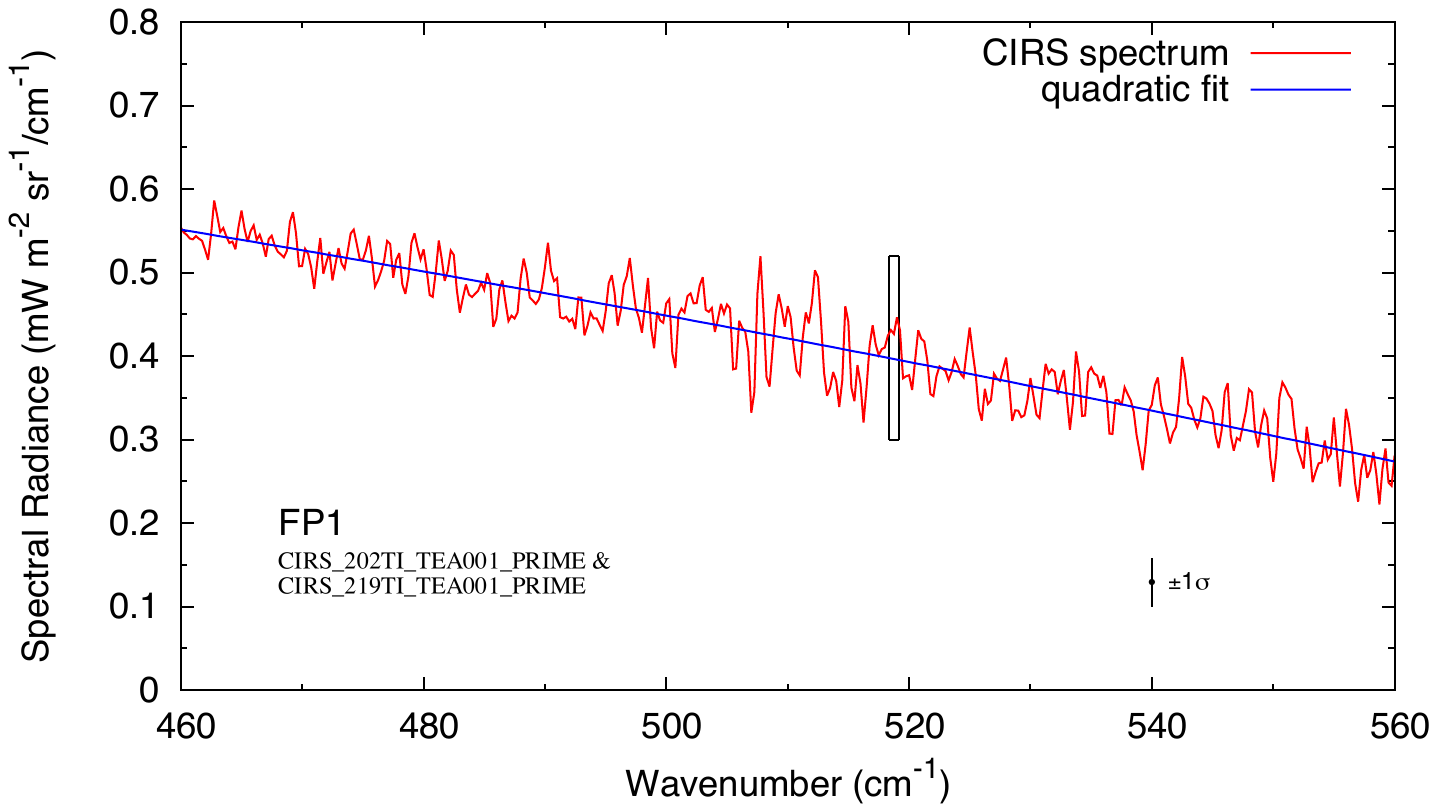}
	  \caption{FP1 Cassini/CIRS spectral average along with a second-order polynomial least-squares fit. The FP1 circular FOV encompasses the whole Titan disk. The narrow rectangle around 519 \cm\ indicates the location and width of the \acd\ TEXES interval. The $\pm$1$\sigma$ noise error bars are indicated.}
	  \label{Fig_obs_CIRS1}
\end{figure}

Regarding the calibration of the \acd\ TEXES setting, we selected FP1 CIRS spectra recorded during a TEA (Titan Exploration at Apoapse) observational sequence designed to place Titan fully within the FP1 pixel \citep{Nixon2019}. Three TEA sequences are useable (CIRS\_182TI\_TEA001\_PRIME, 
CIRS\_202TI\_TEA001\_PRIME and CIRS\_219TI\_TEA001\_PRIME). We chose to average the last two sequences in which the sub-spacecraft latitude is 0 and 51\dg N, the first one being 41\dg S, far from the 27\dg N of our TEXES observations. Details of the CIRS observations are given in Table~\ref{table_CIRS}. The resulting spectrum is shown in Fig.~\ref{Fig_obs_CIRS1} along with a second-order polynomial fit over the range 460--560 \cm\ (a degree 2 for the polynomial fit is the one that minimizes the reduced $\chi ^2$). This fit indicates a continuum intensity of 0.396~$ \pm$~0.002 mW m$^{-2}$ sr$^{-1}$ / cm$^{-1}$ at 519 \cm\ that we need to convert into a flux in Jy at the Titan-Earth distance of the TEXES observations (9.124 AU). 
However, this is not straightforward because the CIRS FP1 FOV is not homogeneous, being greatest at the center, reaching half power at 1.2 mrad from center and zero at 3.2 mrad \citep{Flasar2004, Anderson2011}. To do so, we used the FOV sensitivity function determined by \citet[][Fig.~2]{Anderson2011} and a synthetic spectrum of Titan's emission calculated as a function of distance from disk center. This spectrum was generated from the radiative transfer and atmospheric models described in Section~\ref{analysis}, with the haze opacity adjusted to reproduce the CIRS radiance at 519 \cm . Doing so, we expect a continuum flux of 23.0 Jy in the TEXES observations around this wavenumber. This is about 6\% lower than obtained by using the flux from the STM of 10~Hygiea, and we have then to set the corrective factor $f$ in Eq.~\ref{fluxTexes} to 0.94.

\begin{table}
\caption{Cassini/CIRS Observations of Titan (resolution = 0.5 \cm)}\label{table_CIRS}
\begin{tabular*}{\tblwidth}{@{}LLLLLLL@{}}
\toprule
Sequence  &  Focal & Date & Distance & Number & Sub-solar & Sub-spacecraft  \\ 
&  plane & & (km) & of spectra & latitude & latitude \\ 
\midrule
CIRS\_202TI\_TEA001\_PRIME & FP1 & March 02-03,  & 1,522,600 & 891 & 21 \dg N & 51\dg N \\
&&2014&&&&\\
CIRS\_219TI\_TEA001\_PRIME & FP1 & July 23-24, & 1,580,500 & 784 & 25 \dg N & 0\dg  \\
&&2015&&&&\\
CIRS\_271TI\_COMPMAP001\_ & FP3, & April 23-24,  & 761,800 -- & 3613 (FP3) & 27 \dg N & 17\dg N \\
PRIME & FP4 &2017&917,000&3606 (FP4)&&\\

\bottomrule
\end{tabular*}
\end{table}

\begin{figure}
	\centering
		\includegraphics[angle=0,trim={0 8.3cm 0 0},clip,width=0.9\columnwidth]{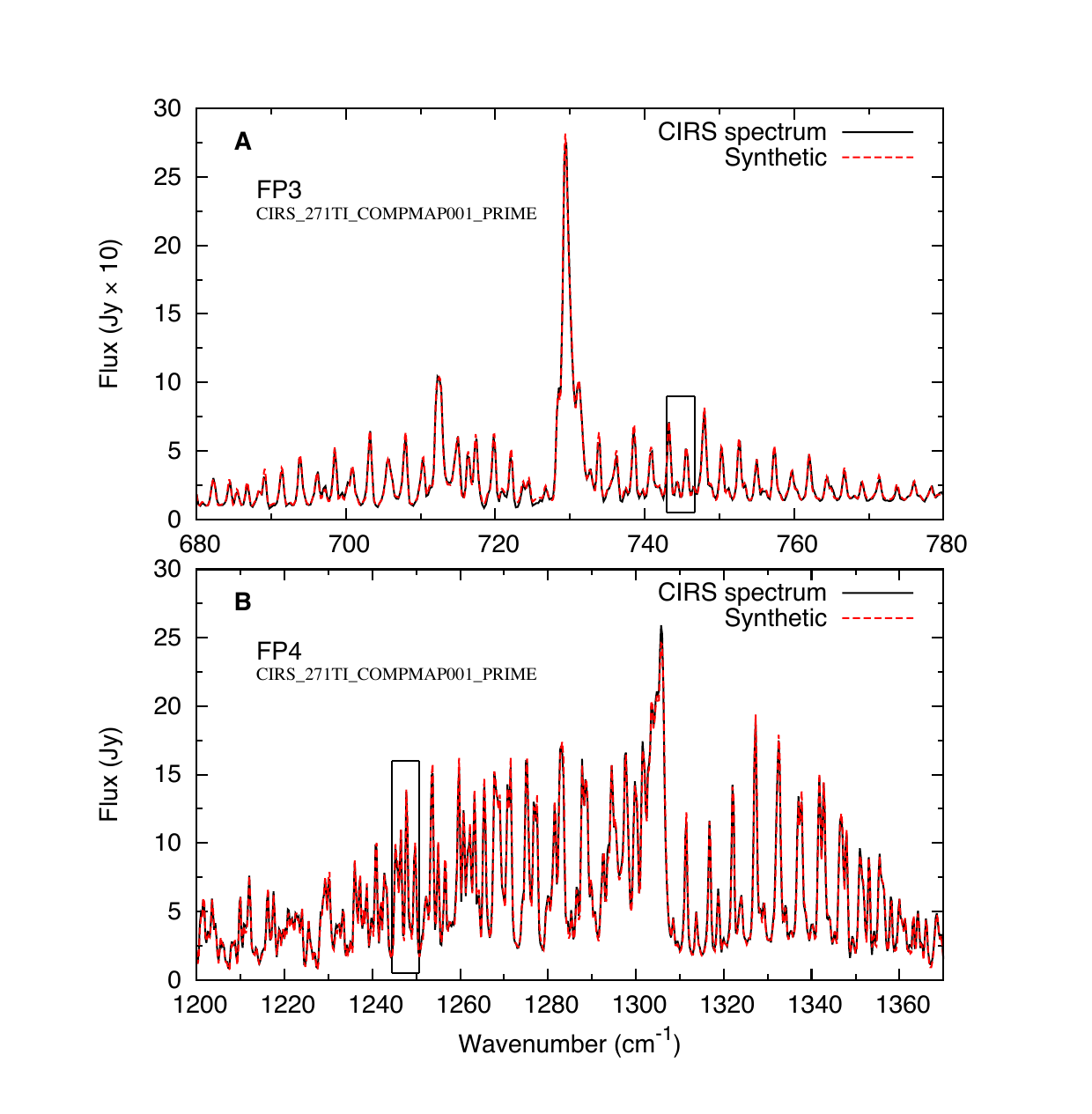}
	  \caption{FP3 (Panel A) and FP4 (Panel B) Cassini/CIRS spectral averages along with best fit models. Fluxes correspond to a Titan--Earth distance of 9.124 AU. The rectangles around 745 \cm\ and 1248 \cm\ indicate the location and width of the \ac\ and \ch\ TEXES intervals respectively. The noise equivalent spectral flux is 0.6 Jy in Panel A ($\sim$0.7 times the line thickness) and 0.10 Jy in Panel B (similar to the line thickness).}
	   \label{Fig_obs_CIRS2}
\end{figure}

The CIRS FP3 and FP4 detectors have a much lower IFOV than FP1's FOV and there are no CIRS observing sequences in which Titan is fully included in their linear arrays. We then turned to the COMPMAP (Composition Map) sequences in which the FP3 and FP4 arrays were positioned to span Titan's disk in one to five positions \citep{Nixon2019}. We selected the one noted in Table~\ref{table_CIRS} as it is closest in terms of time and in sub-solar and sub-instrument latitudes to the TEXES observations. We binned the observations into concentric rings of equal area, centered on Titan's disk and covering altogether 0 to 3050 km in radius (i.e up to 475 km above the surface). We then summed these bins and converted the result into a flux (Jy) at a Titan-Earth distance of 9.124 AU, as relevant for the July 8 TEXES observations. The FP3 and FP4 final spectra are shown in Fig.\ref{Fig_obs_CIRS2}. We degraded the TEXES observations at the resolution of the CIRS spectra (0.53 \cm) with the help of a synthetic spectrum (see Section~\ref{analysis}) to fill in the troughs in the TEXES data. From a least-squares fit approach, we then determined the corrective factor $f$ in Eq.~\ref{fluxTexes} to be 1.09 for the \ac\ setting and 0.86 for the \ch\ setting, taking into account the Titan-Earth distances of the observations (Table~\ref{table_obs}).  

In the radiative transfer analysis of the TEXES observations, we used this CIRS-based flux calibration to which we assign an uncertainty of $\pm$5\%, arising from the construction of the CIRS disk averages and possible variations of the integrated disk flux. We note that the corrective factors we derived vary from 0.86 to 1.09, which seems reasonable given the uncertainty in the 10~Hygiea STM and possible pointing errors.

\section{Radiative transfer analysis}\label{analysis}

Spectral modeling of the observations was performed with a line-by-line radiative transfer code coupled with an iterative inversion scheme \citep{Conrath1998} as described in \cite{Vinatier2007a}. Sources of molecular opacity are the collision-induced opacity (CIA) from N$_2$, \ch\ and H$_2$ pairs, ro-vibrational bands of \ch, CH$_3$D, \ac, \acd, HCN, C$_2$H$_6$ and C$_3$H$_8$, and absorption from aerosol particles. We used the CIA absorption coefficients from \cite{Borysow1986b}, \cite{Karman2015} and  \cite{Finenko2022}. Line positions, energy levels and intensities of \ac\ and \acd\ come from GEISA2020 \citep{Delahaye2021}, the \acd\ parameters being based on the analysis of \cite{Jolly2008}. For the N$_2$ pressure-broadened line widths and temperature exponents, we used the $m$-dependent values determined by \cite{Bouanich1998}. Line parameters for the other molecules are the same as in \cite{Bezard2018}. As regards the aerosol opacity, we used the spectral dependence recommended by \cite{Bezard2020} (their Fig.~4), which is based on \cite{Vinatier2012} beyond 600 \cm. The integrated optical depth at the reference wavenumber of 1090 \cm\ is 0.029.

The atmospheric model includes vertical profiles of gas abundances, aerosol absorption and temperature. The \ch\ profile is fixed to that derived by \cite{Niemann2010} from Huygens in situ measurements. For the other molecular species and the aerosol opacity, we adopted the profiles retrieved by \cite{Vinatier2020} from Cassini/CIRS limb and nadir measurements recorded on September 11, 2017 at 21\dg N. Note that the corresponding \ac\ profile is only used as an a priori (first guess) profile in the retrievals of the \ac\ profile from TEXES measurements. This is also true for the HCN profile. All other absorber abundance profiles are left unchanged in the retrieval code.

The temperature profile that serves as an a priori profile in the inversion process is the 21\dg N profile (September 11, 2017) from \cite{Vinatier2020} down to the 10-mbar level. In the troposphere, the profile is constrained to reproduce two selections of FP1 spectra (15-\cm\ resolution) from 55 to 545 \cm, recorded in 2016--2017 between 0\dg\ and 40\dg N and from a distance of (7.5--9.5)~$\times$ 10$^4$ km from Titan. The first one is characterized by a ``low'' mean emission angle of 13.1\dg\ and the second one a ``high'' mean emission angle of 54.2\dg .
 
Spectra are calculated using an atmospheric grid of 70 layers equally spaced in $\log p$ from the surface up to $p$~= 0.15 $\mu$bar ($\sim$600 km). For Cassini/CIRS modeling, monochromatic spectra are convolved with a Hamming function of resolution 0.528 \cm\ (distance of first zero). For TEXES modeling, the convolution function is a Gaussian with the full width at half-maximum given in Table~\ref{table_obs}. Radiance spectra are calculated for five lines-of-sight intercepting the surface from disk center to the solid radius and for six lines-of-sight above the surface at altitudes up to 475 km. These two regions are divided into respectively five and six circular rings of equal area. The distance of the lines of sight to disk center is taken as the average of the radii of the concentric circles delimiting each annulus. The intensities are then summed with weights equal to the circular ring areas to produce an integrated flux spectrum. 

The inversion algorithm minimizes a weighted sum of the residuals between synthetic and observed spectrum (i.e.\ the $\chi^2$) and between solution and a priori profile (temperature or mole fraction) \citep[see details in][]{Vinatier2007a}.  A correlation length ($L$) of one pressure scale height is further used for filtering the solution profile. The disk-averaged temperature profile in the stratosphere was retrieved from the methane emission spectrum in the 1200--1370 \cm\ range for the CIRS measurements and 1244.7--1250.5 \cm\ for the TEXES measurements, assuming that the stratospheric methane mole fraction is constant with altitude and over the disk \citep[1.48\%;][]{Niemann2010}. In fact, by simultaneously analyzing FP1 and FP4 Cassini/CIRS spectra at different latitudes, \cite{Lellouch2014} showed that the stratospheric methane abundance varies within a range of 1 to 1.5\% as a function of latitude. In particular, the $\sim$1\% mole fraction derived at low latitudes agrees with the reanalysis of the Huygens Descent Imager/Spectral Radiometer (DISR) data by \cite{Rey2018}. The \ch\ abundance profile they derived is lower than the Huygens/GCMS profile \citep{Niemann2010} above $\sim$39 km and reaches $\sim$1\% above 110 km. Consequently, we also ran an alternative case in which we used the \cite{Rey2018} methane profile in place of the \cite{Niemann2010} profile (see Section~\ref{results}).

The disk-averaged acetylene profile was retrieved from the 679--780 \cm\ interval for the CIRS measurements and 742.8--746.6 \cm\ for the TEXES measurements. The \acd~/ \ac\ ratio (assumed to be constant with altitude) was inferred from the TEXES measurements combining the intervals 518.38--518.73 and 519.08--519.17 \cm. The aerosol optical depth at the reference wavenumber was simultaneously retrieved.

The procedure we followed in this analysis is the following:
\begin{enumerate}[a)]
\item retrieve the temperature profile from the disk-averaged CIRS FP4 selection using the \cite{Vinatier2020} 21\dg N profile as a first guess $\longrightarrow T_{\textit{CIRS}}$($p$)
\item retrieve the temperature profile from the TEXES measurements (\ch\ setting and continuum in the \acd\ setting) taking $T_{\textit{CIRS}}$($p$) as the a priori profile $\longrightarrow T_{\textit{TEXES}}$($p$)
\item retrieve the \ac\ profile from the disk-averaged CIRS FP3 selection with the \cite{Vinatier2020}  21\dg N profile for a priori profile and $T_{\textit{CIRS}}$($p$) for the temperature profile $\longrightarrow q_{\textit{CIRS}}$($p$)
\item  retrieve the \ac\ profile from the TEXES measurements (\ac\ setting) using $q_{\textit{CIRS}}$($p$) for the a priori profile and $T_{\textit{TEXES}}$($p$) for the temperature profile $\longrightarrow q_{\textit{TEXES}}$($p$)
\item  determine the \acd~/ \ac\ ratio from the TEXES measurements (\acd\ setting) using $T_{\textit{TEXES}}$($p$) for the temperature profile and $q_{\textit{TEXES}}$($p$) for the \ac\ profile
\end{enumerate}  

We also performed some tests in which we used the Huygens/HASI temperature profile \citep{Fulchignoni2005} recorded in situ at 10\dg S as the a priori profile in Step a or directly in Step b. As regards Step e, we tested an additional case in which we use the CIRS-derived temperature ($T_{\textit{CIRS}}$($p$)) and acetylene ($q_{\textit{CIRS}}$($p$)) profiles to derive the \acd /\ac\ ratio (see next section).

\begin{figure}
	\centering
		\includegraphics[angle=0,trim={0 5cm 0 2cm},clip,width=0.9\columnwidth]{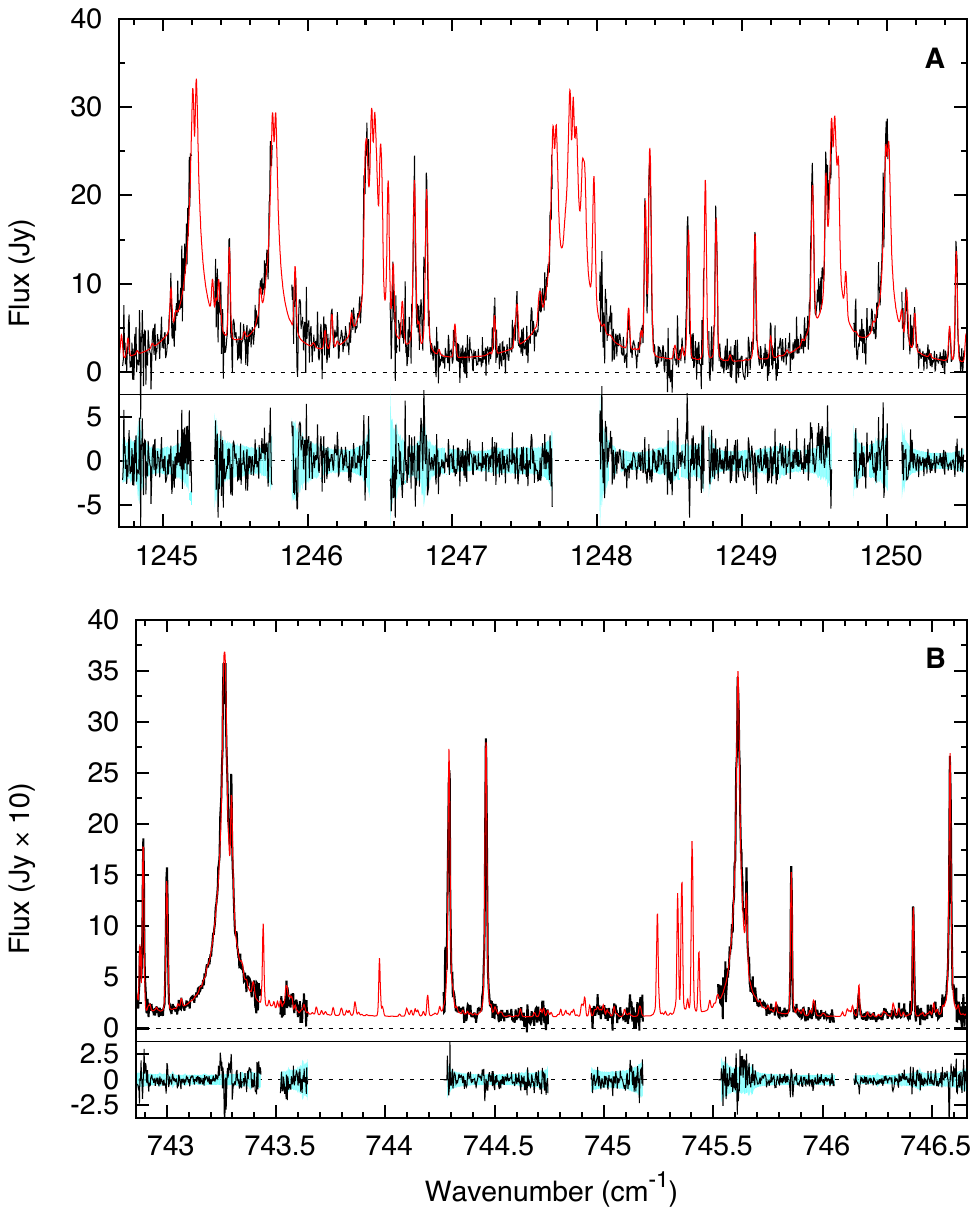}
	  \caption{Best model fit (red) to observed TEXES (black) spectrum in the \ch\ setting around 1248 \cm\ (Panel A) and in the \ac\ setting around 745 \cm\ (Panel B). The lower sub-panels in both panels show the residuals of the fit (black) along with the $\pm$1$\sigma$ noise envelope (cyan). The TEXES spectra are calibrated against Cassini/CIRS spectral averages (see Section~\ref{CIRS_obs})}
	   \label{Fig_fit_Texes1}
\end{figure}

\section{Results}\label{results}
\begin{figure}
	\centering
		\includegraphics[angle=-90,trim={0.5cm 1.5cm 0.8cm 0},clip,width=1\columnwidth]{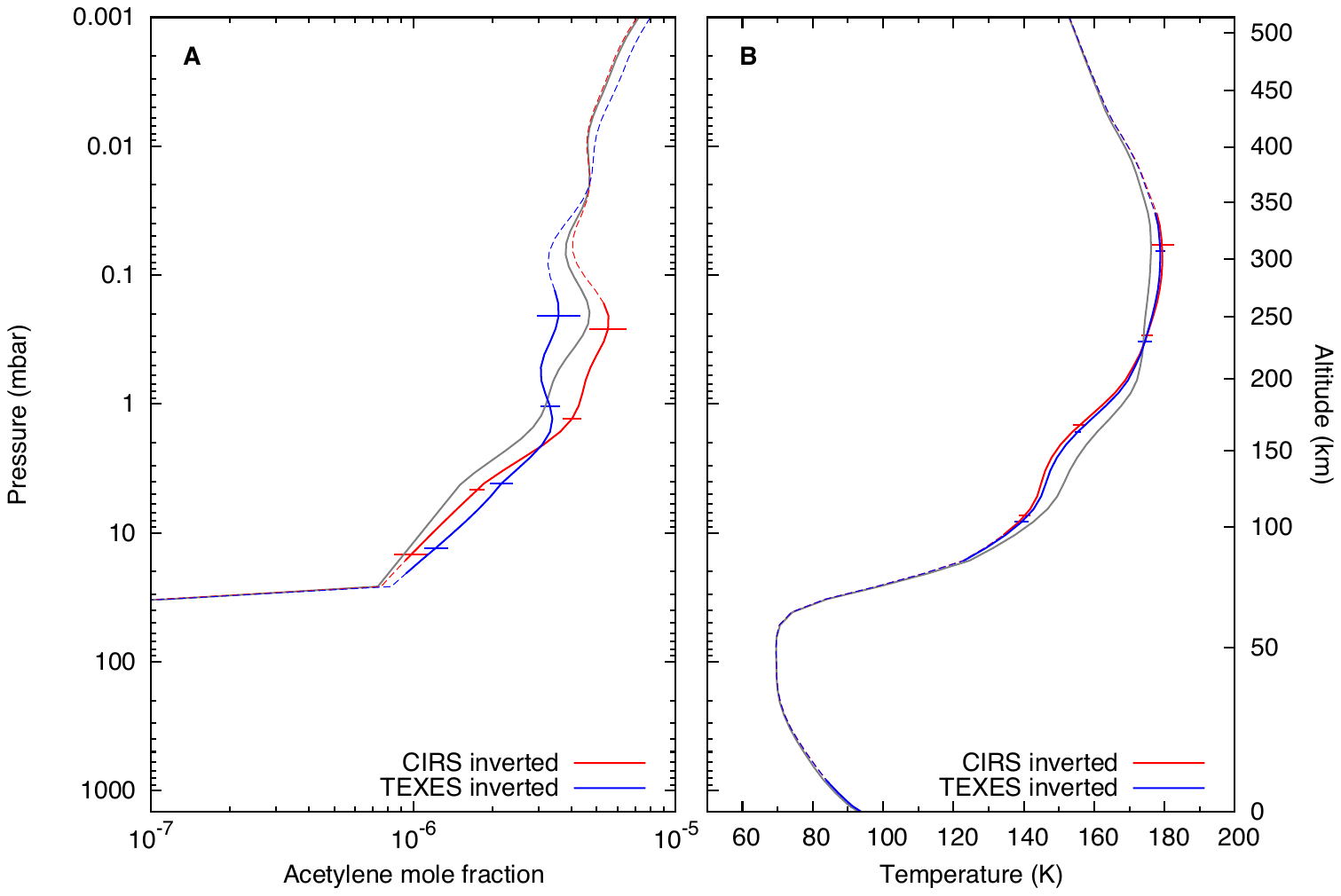}
	  \caption{Vertical profiles of \ac\ (Panel A) and temperature (Panel B) retrieved from the Cassini/CIRS (red) and TEXES data (blue). A stratospheric methane abundance of 1.48\% was used in the temperature retrievals. The gray profiles are the a priori \cite{Vinatier2020} profiles used in the inversion of the CIRS spectral averages. The solid colored lines correspond to the regions to which the data are sensitive. The formal 1--SD noise errors on the TEXES-retrieved profiles and the regions probed by the observations are: 0.2 K (0.03--15 mbar) for the temperature and 2\% for the \ac\ mole fraction (0.10--20 mbar). The total error bars, including those from model uncertainties, are indicated at a few selected levels}
	   \label{Tq_profiles}
\end{figure}

\begin{figure}
	\centering
		\includegraphics[angle=-90,trim={0.5cm 1.5cm 0.8cm 0},clip,width=1\columnwidth]{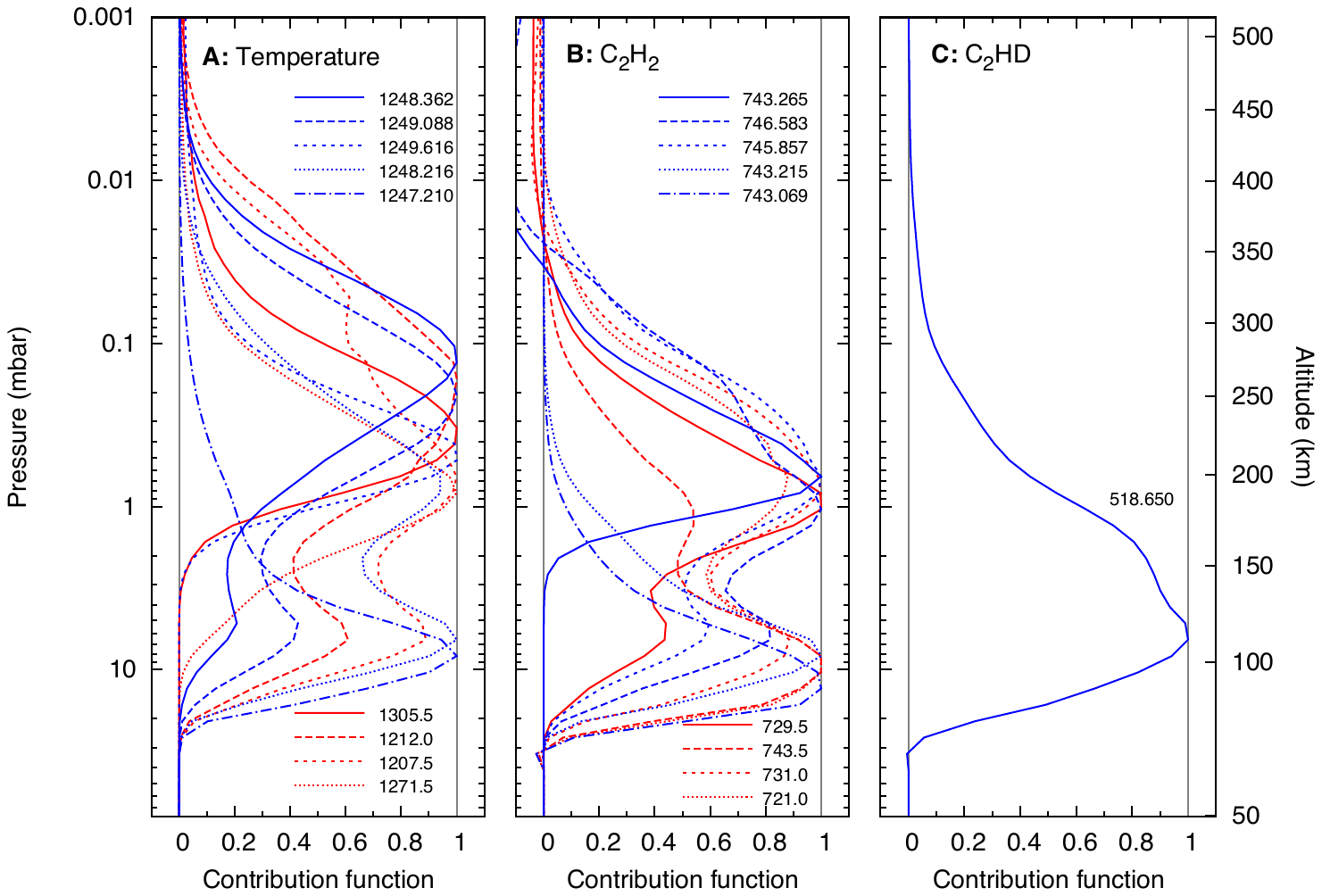}
	  \caption{Contribution functions at selected wavenumbers calculated for the best-fitting models (Fig.~ \ref{Tq_profiles}) of CIRS (red lines) and TEXES (blue lines) observations. These contribution functions, which have been normalized at each wavenumber, are Jacobians that give the rate of change of spectral flux with temperature (Panel A), logarithm of \ac\ mole fraction (Panel B), and logarithm of \acd\ mole fraction (Panel C).}
	   \label{CF}
\end{figure}

The temperature profile retrieved from the inversion of the CIRS FP4 average ($T_{\textit{CIRS}}$($p$)) is shown in Fig.~\ref{Tq_profiles} and the corresponding model fit to the data is shown in Fig.~\ref{Fig_obs_CIRS2}b. The observations are sensitive to the 0.03--11 mbar range (95--350 km) (Fig.~\ref{CF}). In the 0.1-mbar region, this profile is warmer than the a priori profile, which refers to latitudes around 21\dg N, while the situation is reversed in the 1--10 mbar range. This is expected as the CIRS spectral average incorporates latitudes that are on the average warmer than mid-northern latitudes around 0.1 mbar and cooler in the 1--10 mbar pressure range, as can be seen in Fig.~2 of \cite{Vinatier2020}.

In a second step, this CIRS-inverted profile was used as an a priori profile to retrieve a temperature profile $T_{\textit{TEXES}}$($p$) from the TEXES observations around 1248 \cm. This profile is shown in Fig.~\ref{Tq_profiles} and the corresponding fit of the data in Fig.~\ref{Fig_fit_Texes1}a. The higher spectral resolution of the TEXES data allows us to somewhat tighten the constraints in the lower stratosphere down to about 15 mbar (85 km) (Fig.~\ref{CF}). In the inversion procedure, we also used data in small spectral intervals in the continuum around 518.5--518.7 \cm\ to constrain the (near-)surface temperature. The CIRS-- and TEXES--retrieved temperature profiles are very similar and differ by at most 1.5~K in the 2--mbar region. 

A \ac\ profile $q_{\textit{CIRS}}$($p$) was first obtained from the CIRS FP3 spectral average using the 21\dg N profile of \cite{Vinatier2020} as the a priori in the inversion process (Fig.~\ref{Tq_profiles}). The fit to the data is shown in Fig.~\ref{Fig_obs_CIRS2}a. The region probed ranges from 0.12 mbar (275 km) to 18 mbar (80 km) (Fig.~\ref{CF}). The TEXES data around 745 \cm\ were then used to retrieve the final \ac\ profile ($q_{\textit{TEXES}}$($p$)) shown in blue in Fig.~\ref{Tq_profiles} using the CIRS-retrieved profile as the a priori. The fit of the data and the residuals are shown in Fig.~\ref{Fig_fit_Texes1}b. As for temperature, the high resolution of the TEXES observations allows us to slightly improve the constraints on the disk-averaged \ac\ profile, extending the region probed to $\sim$0.10--20 mbar (75--285 km). The TEXES-derived \ac\ mole fraction is some 35\% smaller than the CIRS-derived one in the 0.2--0.5 mbar range and about 20\% larger from 4 to 15 mbar.

\begin{figure}
	\centering
		\includegraphics[angle=0,trim={0 9.5cm 0 9cm},clip,width=0.9\columnwidth]{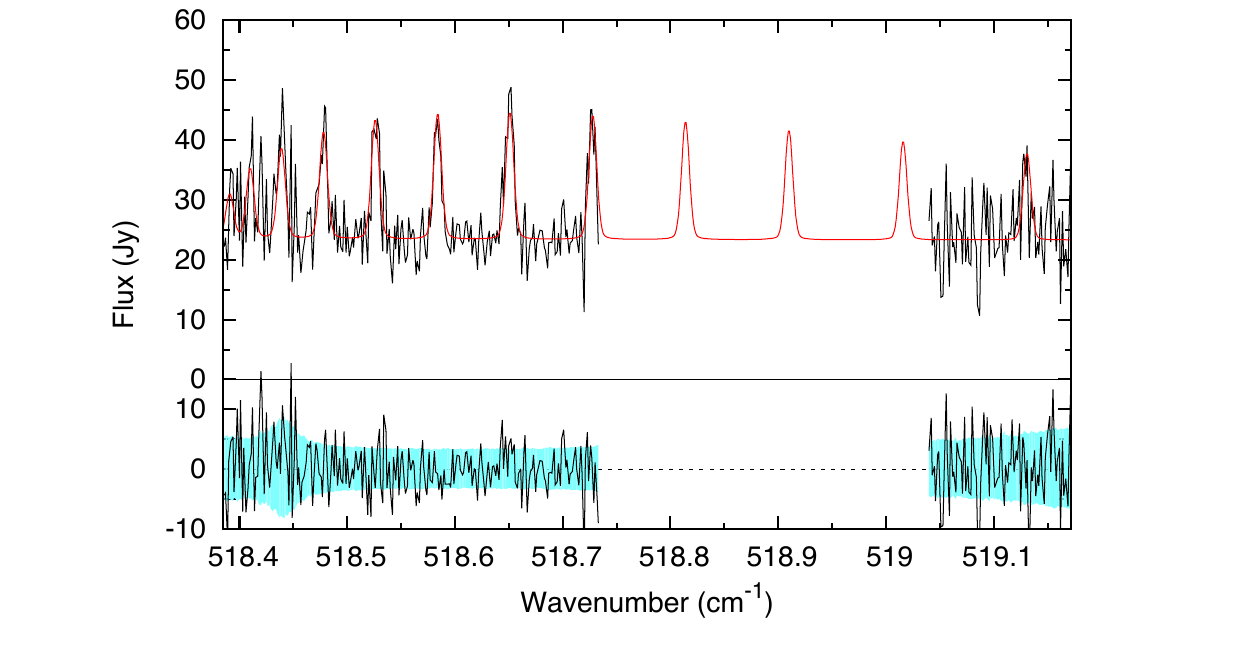}
	  \caption{Best model fit (red) to observed TEXES (black) spectrum in the \acd\ setting around 519 \cm. The lower sub-panel shows the residuals of the fit (black) along with the $\pm$1$\sigma$ noise envelope (cyan). The TEXES spectrum is calibrated against a Cassini/CIRS spectral average (see Section~\ref{CIRS_obs}).}
	   \label{Fig_fit_Texes2}
\end{figure}

The formal 1--SD noise errors on the TEXES-retrieved profiles are 0.2 K in the 0.03--15 mbar range for the temperature and 2\% for the \ac\ mole fraction in the 0.10--20 mbar. We evaluated the model errors from the variations induced in the retrieved profiles by a change in the a priori temperature and \ac\ abundance profiles used in Steps a) and c). To do so, we considered a priori profiles that differ from the \cite{Vinatier2020} 21\dg N profiles above the 20-mbar level by $\pm$10~K (temperature) and factors of 0.5 or 2 (\ac). The so-derived error bars are shown in Fig.~\ref{Tq_profiles} at a few pressure levels.

Finally, we used the TEXES \acd\ setting around 519 \cm\ to retrieve the \acd~/ \ac\ ratio using the TEXES-derived temperature profile $T_{\textit{TEXES}}$($p$) and \ac\ profile $q_{\textit{TEXES}}$($p$). In the inversion process, the \acd~/ \ac\ ratio is assumed to be constant with altitude and we thus solved for only one parameter (technically this is done by fixing a very large value for the correlation length $L$, 100 times the pressure scale height). We derived a \acd/\ac\ ratio that corresponds to a D/H ratio of (1.14$\pm$0.04)~$ \times$ 10$^{-4}$, where the error bars account only for the noise equivalent flux in the TEXES \acd\ observations and correspond to a $\chi^2$ variation of 1.  The model fit to the observations is shown in Fig.~\ref{Fig_fit_Texes2}. Our measurement is most sensitive to the region between 0.7 and 16 mbar (Full Width at Half Maximum of the contribution function), with the contribution function peaking at 6 mbar (115 km) (Fig.~\ref{CF}).

We investigated different sources of error. We made a first test in which we used the CIRS-derived atmospheric model ($T_{\textit{CIRS}}$($p$) and $q_{\textit{CIRS}}$($p$)) to model the \acd\ TEXES lines. We then derived a D/H ratio of (1.09$\pm$0.04)~$\times$ 10$^{-4}$, i.e. 5\% smaller than when the temperature and acetylene profiles are derived from the \ch\ and \ac\ TEXES settings. We also tested the sensitivity of the derived D/H ratio to assumptions about the a priori profiles used in the retrieval process. First we used the Huygens/HASI profile \citep{Fulchignoni2005} as the a priori profile for the temperature retrievals from Cassini/CIRS data (Step a) or directly for the temperature retrievals from TEXES data (Step b). In either case, the so-derived D/H ratio is within 1\% of our nominal value. Secondly, we considered the propagation of the model errors mentioned above into the retrieved D/H ratio. A variation of the a priori temperature model by $\pm$10~K induces a change by $\pm$3\% in the derived D/H ratio while a variation of the a priori acetylene model by a factor of two causes negligible variations of the retrieved D/H ratio ($<$1\%).

As discussed in Section~\ref{analysis}, we also ran a case in which the temperature was retrieved using the \cite{Rey2018} methane profile, which has a stratospheric \ch\ abundance of 1\% instead of the 1.48\% mole fraction from \cite{Niemann2010}. In this case, the derived temperature profile is warmer by at most 2.6~K at 1.3 mbar and by 1.2~K at 0.25 and 10 mbar. The derived \ac\ mole fraction is consequently lower by up to -32\% around 1 mbar while the D/H ratio reaches 1.31~$ \times$ 10$^{-4}$, the \acd\ lines being less sensitive to temperature change than the optically thick \ac\ lines. Because the methane mole fraction varies over Titan's disk between 1 and 1.5\% \citep{Lellouch2014}, we retained for the nominal value of the D/H ratio the (geometrical) mean of the values retrieved for the \cite{Niemann2010} and for the \cite{Rey2018} \ch\ profiles, these values providing the conservative upper and lower limits: D/H (in \ac) = (1.22$^{+0.09}_{-0.08}$) $\times$ 10$^{-4}$.

We also considered a 5\% flux calibration error in each of the three TEXES settings and finally took into account the $\pm$4\% uncertainty of the \acd\ band intensity reported by \cite{Jolly2008}. The errors induced on the D/H ratio are listed in Table~\ref{error}. Combining quadratically these different error sources, we obtain a relative error of +24\% / -18\%, leading to:
\begin{center}
D/H (in \ac) = (1.22$^{+0.27}_{-0.21}$)~$\times$ 10$^{-4}$
\end{center}

We note that an important source of error is that due to the flux calibration of the \ac\ setting. A 5\% variation of the flux scale causes a change of the retrieved \ac\ abundance profile of 24\% at 1mbar, 11\% at 5 mbar and 6\% at 10 mbar. The induced change on the \acd~/ \ac\ ratio is then +17\% / -12\%. The second most significant source of error is the uncertainty on the disk-averaged stratospheric \ch\ used in the temperature retrievals.

\begin{table}
\caption{Relative uncertainties on the derived D/H ratio in acetylene}\label{error}
\begin{tabular*}{0.33\textwidth}{@{}LL@{}}
\toprule
Origin  &  Relative error  \\ 

\midrule

Noise (1$\sigma$) & $\pm$3.5\%  \\
Use CIRS $T$ and $q$ & -5\% \\
$T$ and $q$ model errors & $\pm$3\% \\
\ch\ abundance & +8\% / -7\% \\
5\% flux error in: &   \\
 -- \acd\ setting &  $\pm$6\% \\
 -- \ac\ setting &  +17\% / -12\% \\
 -- \ch\ setting &  $\pm$5.5\% \\
 Band intensity & $\pm$4\% \\
 \midrule
 Total &  +22\% / -17\% \\

\bottomrule
\end{tabular*}
\end{table}

\section{Discussion}\label{discussion}

The value we retrieved for the D/H ratio in acetylene (1.22$^{+0.27}_{-0.21}$ $\times$ 10$^{-4}$) is significantly smaller than that derived by \cite{Coustenis2008} from different selections of Cassini/CIRS nadir spectra over Titan's disk (2.09$\pm$0.45 $\times$ 10$^{-4}$). It is also marginally smaller than the value derived by the same authors from Cassini/CIRS limb data at four different latitudes, averaging to 1.63$\pm$0.27 $\times$ 10$^{-4}$, although their error bars clearly overlap with ours. These CIRS determinations are made difficult by the weakness of the \acd\ emission in the 678-\cm\ $\nu _5$ band at the instrument's resolution of 0.5 \cm. In addition, the part of the analysis limited to surface-intercepting measurements suffers from the lack of a precise determination of the \ac\ vertical profile from the data, which has a strong influence on the retrieved \acd\ / \ac\ ratio. On the other hand, the precision of our measurement is mostly limited by the fact that we do not observe the \acd\ and \ac\ lines simultaneously. We have then to rely on a relative calibration of the two instrumental settings for which we used Cassini/CIRS spectral averages as representative as possible of the observing conditions of TEXES (period, sub-solar and sub-instrument latitudes).

The D/H ratio in acetylene we obtained is consistent, within error bars, with the D/H ratio measured in methane from Cassini/CIRS measurements by \cite{Bezard2007} (1.32$^{+0.15}_{-0.11}$ $\times$ 10$^{-4}$), \cite{Coustenis2007} (1.17$^{+0.23}_{-0.28}$ $\times$ 10$^{-4}$), \cite{Abbas2010} (1.58$\pm {0.16}$ $\times$ 10$^{-4}$) and \cite{Nixon2012} (1.59$\pm {0.27}$ $\times$ 10$^{-4}$) and from the ground-based measurements by
\cite{Penteado2005} (1.25$\pm {0.25}$ $\times$ 10$^{-4}$) and \cite{deBergh2012} (1.13$\pm {0.25}$ $\times$ 10$^{-4}$), indicating altogether a value in the range (1.1-1.6) $\times$ 10$^{-4}$. 

This result suggests no significant fractionation of deuterium in acetylene compared with methane, the main atmospheric reservoir of hydrogen. On the other hand, as first investigated by \cite{Pinto1986} and then \cite{Lunine1999}, a source of fractionation in the methane photochemistry lies in the higher energy of the C--D bond compared with the C--H bond. These authors considered that methane is most efficiently destroyed by the abstraction reactions: \\
C$_2$H + CH$_4$  $\rightarrow$ \ac\ + CH$_3$ \hfill (k$_1$)\\
C$_2$H + CH$_3$D  $\rightarrow$ \ac\ + CH$_2$D \hfill (k$_{2a}$)\\
\hphantom{C$_2$H + CH$_3$D}  $\rightarrow$ C$_2$HD + CH$_3$ \hfill (k$_{2b}$)\\
From consideration of chemical kinetics of some deuterated species, \cite{Lunine1999} argued that $q$, equal to ($k_{2a} + k_{2b}$)/$k_1$, lies in the range 0.80--0.88. Because the ratio of the reaction coefficients $q$ is lower than 1, \ch\ destruction is favored over CH$_3$D destruction, opening up the possibility of a deuterium fractionation in the photochemical products. We define $q'$ as the ratio of the intrinsic probability of breaking a C-D bond to that of breaking a C-H bond, or equivalently by the ratio $3 k_{2b}/k_{2a}$. If we further assume that $k_{2a}$ is equal to 0.75 $k_1$, given that Reaction 2a does not break the C--D bond, then $q'$ =$3 k_{2b}/k_{2a}$ =  $4 k_{2b}/k_{1}$ is in the range 0.2--0.52.

According to the photochemical model of \cite{Vuitton2019}, photolysis of ethylene (\ethy) is the main source of acetylene in the bulk of the atmosphere. This production peaks around 300 km. Ethylene is mostly produced in the upper atmosphere ($\sim$800 km) from the reaction of the CH radical with methane and flows down to the stratosphere where it is mostly lost by photolysis. In Appendix~\ref{photochemistry}, we use a simplified 0-D photochemical model to identify possible fractionation processes in the production of \ac\ linked to the expected higher binding energy of the C--D bond. We found that fractionation of deuterium can occur at two stages: in the formation of the CH radical from the $^3$CH$_2$ radical and from the photolysis of \ethy. Both of them tend to enhance the D/H ratio in \ethy\ and thus in \ac. If the $q'$ ratios associated to these two reactions were similar to that for C$_2$H + CH$_4$ mentioned above (0.2--0.52), the D/H in \ac\ would be as high as 1.4--1.9 times that in \ch\ (Eq.~\ref{f_C2H2_tot}), which is not what we observe. On the other hand, if fractionation were limited to the CH radical formation stage (i.e., assuming no fractionation through the \ethy\ photolysis), the fractionation factor would be lower (1.08--1.17) and compatible with the D/H measurement errors.This implies that either the kinetic isotope effect (KIE) in the photodissociation of \ethy\ is not as strong as in the C$_2$H + CH$_4$ reaction considered by \cite{Lunine1999}, or that other photochemical processes that we did not consider in Appendix~\ref{photochemistry} mitigate this KIE.

\section{Conclusions}\label{conclusions}

We report here the first observation of several lines from the Q-branch of the $\nu_4$ band of \acd\  on Titan  at 19.3 $\mu$m. Combining observations of \acd, \ac\ and \ch\ lines with the same instrument (IRTF/TEXES) and using Cassini measurements for the flux calibration, we derived a value for the D/H ratio in acetylene of (1.22$^{+0.27}_{-0.21}$) $\times$ 10$^{-4}$. This value is consistent within error bars with the D/H ratio measured in methane. Our main source of uncertainty arises from the relative flux calibration of the three different settings we used to record the \acd, \ac\ and \ch\ lines. Improving on the flux calibration, e.g. by observing these emissions simultaneously, would allow us to reduce significantly the error bars on the D/H ratio in acetylene. Recent observations by the MIRI medium-resolution spectrometer (MRS) aboard the James Webb Space Telescope (JWST) conducted in July 2023 (Titan project GTO 1251) may help in this regard once issues in fringe removal and flux calibration have been resolved.

While our result indicates no significant deuterium fractionation in acetylene relative to methane, at least in the region probed (0.7-16 mbar; 80-190 km), we have identified two possible processes in the photochemical production of \ac\ that should in theory enhance the D/H ratio in acetylene. These are the formation of the CH radical from the $^3$CH$_2$ radical and the photodissociation of \ethy. This potential enhancement arises from the fact that the C--D bond is slightly stronger than the C--H bond. Laboratory measurements of the KIEs for these two reactions, most importantly for the \ethy\ photolysis, would be very valuable. It would be also useful that complete photochemical models incorporate deuterium species, even with approximative KIEs, to investigate the deuterium fractionation in various hydrocarbons. From an observational standpoint, signatures from other deuterated hydrocarbons (e.g.\ C$_2$H$_6$, \ethy) could also be sought in Titan's spectrum to better constrain its atmospheric photochemistry.

\section*{Data availability}

The raw TEXES data used in this study can be obtained from the NASA/IRTF archive at \url{https://irsa.ipac.caltech.edu/applications/irtf/}. The product IDs for the three spectral settings (\acd, \ac\ and \ch) are:
\begin{description}
\item[Titan C$_2$HD:] TX17A0708.1038-1045
\item[10~Hygiea C$_2$HD:] TX17A0708.1035-1036
\item[Titan C$_2$H$_2$:] TX17A0714.6004-6017
\item[10~Hygiea C$_2$H$_2$:] TX17A0714.6018-6019
\item[Titan CH$_4$:] TX17A0708.1030-1033
\item[10~Hygiea CH$_4$:] TX17A0708.1017-1028
\end{description}
The raw data were calibrated using the 2017 version of the TEXES pipeline software (current version is available at \url{https://github.com/TEXESArch/pipecode}). The reduced data presented in this study are archived \citep{Bezard2024} as ascii files.

\section*{Acknowledgments}

BB, SV and EL acknowledge support from the Programme National de Plan\'etologie (PNP) of CNRS-INSU co-funded by CNES. TK was a Visiting Astronomer at the Infrared Telescope Facility, which is operated by the University of Hawaii under contract 80HQTR19D0030 with the National Aeronautics and Space Administration.



\appendix
\section{Fractionation in the production of \ac }\label{photochemistry}

Potential sources of isotopic fractionation in the photochemistry of methane include photochemical reactions (molecular, ion-molecule and photodissociation), atmospheric escape, condensation/evaporation and molecular diffusion. We investigate here how the photochemical production of \ac\ from methane photochemistry could lead to a deuterium fractionation due to the greater strength of the C--D bond compared to the C--H bond. To do so, we only considered the major chemical steps leading to the formation and loss of \ac\ (through 0-D calculations), based on the photochemical model of \cite{Vuitton2019}.  We further assumed that a reaction of a species A (radical or molecule) with a deuterated hydrocarbon B has the same reaction rate $k$ as with the main isotopologue if it does not break the C--D bond and a rate $q' k$, with 0$\leq q' \leq$1, if the C--D bond is broken. The same is assumed for a photodissociation reaction.

According to \citeauthor{Vuitton2019}'s (\citeyear{Vuitton2019}) model, \ac\ is mostly formed by photodissociation of \ethy\ in the upper stratosphere ($\sim$300 km):\\
\ethy\ + $h\nu$  $\rightarrow$ \ac\ + 2H \hfill (J$_1$)\\
\hphantom{\ethy\ + $h\nu$}  $\rightarrow$ \ac\ + H$_2$, \hfill (J$_2$)\\
and accordingly for \acd\ by:\\
\ethyd\ + $h\nu$  $\rightarrow$ \acd\ + 2H \hfill (0.5 J$_1$)\\
\hphantom{\ethyd\ + $h\nu$}  $\rightarrow$ \acd\ + H$_2$ \hfill (0.5 J$_2$)\\
Note that in addition, \ethyd\ can be dissociated as:\\
\ethyd\ + $h\nu$  $\rightarrow$ \ac\ + H + D \hfill (0.5 $q'_d$ J$_1$)\\
\hphantom{\ethyd\ + $h\nu$}  $\rightarrow$ \acd\ + HD \hfill (0.5 $q'_d$ J$_2$)\\

\ac\ and \acd\ are both lost by vertical transport to the condensation region in the lower stratosphere ($\sim$70 km), with a loss rate proportional to the species number density in this formation region: $K_1$ [\ac] and $K_1$ [\acd]. We neglect here the small vapor pressure isotope effect in the relative condensation loss of the two isotopologues. Equaling the production and loss rates for both isotopologues yields:
[\ac] = J$_1$ [\ethy] / $K_1$ and [\acd] = 0.5 J$_1$ [\ethyd] / $K_1$, so that 
\begin{equation}\label{f_C2H2}
\mathrm{[D/H]}_{C_2H_2} = 0.5\,  \mathrm{[ C_2HD]/[ C_2H_2]} = 0.25\,  \mathrm{[C_2H_3D]/[C_2H_4]} = \mathrm{[D/H]}_{C_2H_4}^1,
\end{equation}
where the latter term is the D/H ratio in ethylene in the formation region of acetylene (upper stratosphere).

Ethylene is transported from its formation region in the thermosphere where it has a D/H ratio that we denote [D/H]$_{C_2H_4}^0$. The \ethy\ used to produce \ac\ is therefore supplied by a source having this D/H ratio and lost by photolysis with a \ethyd\ to \ethy\ loss ratio equal to  0.5 (1+$q'_d$) J [\ethyd] / (J [\ethy] ) = 0.5 (1+$q'_d$) [\ethyd] /  [\ethy], with J = J$_1$ + J$_2$. Therefore,  in steady state and neglecting other sinks for ethylene, we obtain:
\begin{equation}\label{f_C2H4_loss}
\mathrm{[D/H]}_{C_2H_4}^1 = 2/(1+q'_d)\, \mathrm{[D/H]}_{C_2H_4}^0
\end{equation}
Photodissociation of ethylene thus appears as a potential source of enhancement of the D/H ratio in this species and consequently in acetylene, provided that $q'_d <$1.

Following \cite{Vuitton2019}, we assume that the bulk of ethylene is formed in the thermosphere, around 800 km, through the reaction:\\
CH + \ch\ $\rightarrow$ \ethy\ + H,\hfill (k$_1$)\\
while \ethyd\ is accordingly formed by:\\
CH + \chd\ $\rightarrow$ \ethyd\ + H, and\hfill (0.75 k$_1$)\\
CD + \ch\ $\rightarrow$ \ethyd\ + H\hfill (k$_1$)\\
The CH radical is mostly formed from the suite of reactions:\\
\ch\ + $h\nu$  $\rightarrow$ $^1$CH$_2$ + 2H\hfill (J$_3$)\\
$^1$CH$_2$ $\rightarrow$ $^3$CH$_2$\\
$^3$CH$_2$ + H $\rightarrow$ CH + H$_2$\hfill (k$_2$),\\
where $^3$CH$_2$  is the methylene radical in the ground electronic state X$^3$B$_1$ and $^1$CH$_2$ the methylene radical in the excited state a$^1$A$_1$.
The CHD radical is formed through:\\
\chd\ + $h\nu$ $\rightarrow$ $^1$CHD + 2H\hfill (0.5 J$_3$)\\
and lost through:\\
$^3$CHD + H $\rightarrow$ CD + H$_2$\hfill (0.5 k$_2$)\\
\hphantom{$^3$CHD + H} $\rightarrow$ CH + HD\hfill (0.5 $q'_f$ k$_2$)\\
The former reaction also forms CD, which is lost through the CD + \ch\ reaction mentioned above.

Balancing production and loss rates for each radical and assuming [CH$_3$D] $<<$ [CH$_4$] yields:\newline
[$^3$CH$_2$] = J$_3$ [\ch] / (k$_2$ [H]),\newline
[CH] = J$_3$ / k$_1$,\newline
[$^3$CHD] = J$_3$ [\chd] / (k$_2$ (1 + $q'_f$) [H]), and\newline
[CD] = (0.5 / (1 + $q'_f$)) J$_3$ [\chd] / (k$_1$ [\ch]).

Ethylene is mostly lost by downward transport, a process that is not isotope-selective and proportional to the number density. Denoting the loss rates for the two isotopologues as $K_0$ [\ethy] and $K_0$ [\ethyd] and again balancing production and loss rates, we obtain:\newline
[\ethy] = J$_3$ [\ch] / $K_0$,\newline
[\ethyd] = J$_3$ [\chd] (0.75 + 0.5 / (1 + $q'_f$)) / $K_0$

The D/H ratio in ethylene in its production region is then given by:
\begin{equation}\label{f_C2H4_prod}
\mathrm{[D/H]}_{C_2H_4}^0 = (0.75 + 0.5/(1+q'_f))\, \mathrm{[D/H]}_{CH_4}
\end{equation}

A fractionation of deuterium may then occur in the production of ethylene, linked to the reaction of the methylene radical (CH$_2$) with the H atom. Combining Eqs.~\ref{f_C2H2}--\ref{f_C2H4_prod}, the D/H ratio in acetylene in this simplified model is equal to: 
\begin{equation}\label{f_C2H2_tot}
\mathrm{[D/H]}_{C_2H_2} = 2 \, (0.75 + 0.5/(1+q'_f)) / (1 + q'_d) \, \mathrm{[D/H]}_{CH_4}
\end{equation}

If the $q'$ factors associated to \ethy\ photodissociation ($q'_d$) and to reaction of CH$_2$ with H ($q'_f$) were in the range of that for the C$_2$H + CH$_4$ reaction \citep[$q'$ = 0.2--0.52 according to][]{Lunine1999}, the D/H ratio in acetylene would be enhanced by a factor in the range 1.42--1.94 compared to the ratio in methane. This is not consistent with our measurement. On the other hand, if fractionation were only effective for the CH$_2$ + H reaction and not for the \ethy\ photodissociation ($q'_d$ = 1), the fractionation factor would amount to 1.08--1.17, which is compatible with our result, given the errors associated with D/H ratio measurements in acetylene and methane. The kinetic isotope effect (KIE) in \ethy\ photodissociation appears to be potentially the most important factor in the deuterium fractionation in acetylene. Our result of D/H in \ac\  being so similar to that in \ch\ suggests that the D/H fractionation in \ac\ via the photolysis of \ethy\ should be weaker than in the catalytic destruction of methane through the C$_2$H + CH$_4$ reaction.

\printcredits

\bibliographystyle{cas-model2-names}

\bibliography{Titan}

\begin{thebibliography}{56}
\expandafter\ifx\csname natexlab\endcsname\relax\def\natexlab#1{#1}\fi
\providecommand{\url}[1]{\texttt{#1}}
\providecommand{\href}[2]{#2}
\providecommand{\path}[1]{#1}
\providecommand{\DOIprefix}{doi:}
\providecommand{\ArXivprefix}{arXiv:}
\providecommand{\URLprefix}{URL: }
\providecommand{\Pubmedprefix}{pmid:}
\providecommand{\doi}[1]{\href{http://dx.doi.org/#1}{\path{#1}}}
\providecommand{\Pubmed}[1]{\href{pmid:#1}{\path{#1}}}
\providecommand{\bibinfo}[2]{#2}
\ifx\xfnm\relax \def\xfnm[#1]{\unskip,\space#1}\fi
\bibitem[{{Abbas} et~al.(2010){Abbas}, {Kandadi}, {LeClair}, {Achterberg},
  {Flasar}, {Kunde}, {Conrath}, {Bjoraker}, {Brasunas}, {Carlson}, {Jennings}
  and {Segura}}]{Abbas2010}
\bibinfo{author}{{Abbas}, M.M.}, \bibinfo{author}{{Kandadi}, H.},
  \bibinfo{author}{{LeClair}, A.}, \bibinfo{author}{{Achterberg}, R.K.},
  \bibinfo{author}{{Flasar}, F.M.}, \bibinfo{author}{{Kunde}, V.G.},
  \bibinfo{author}{{Conrath}, B.J.}, \bibinfo{author}{{Bjoraker}, G.},
  \bibinfo{author}{{Brasunas}, J.}, \bibinfo{author}{{Carlson}, R.},
  \bibinfo{author}{{Jennings}, D.E.}, \bibinfo{author}{{Segura}, M.},
  \bibinfo{year}{2010}.
\newblock \bibinfo{title}{{D/H ratio of Titan from Observations of the
  Cassini/Composite Infrared Spectrometer}}.
\newblock \bibinfo{journal}{Astrophys. J.} \bibinfo{volume}{708},
  \bibinfo{pages}{342--353}.
\newblock \DOIprefix\doi{10.1088/0004-637X/708/1/342}.
\bibitem[{{Anderson} and {Samuelson}(2011)}]{Anderson2011}
\bibinfo{author}{{Anderson}, C.M.}, \bibinfo{author}{{Samuelson}, R.E.},
  \bibinfo{year}{2011}.
\newblock \bibinfo{title}{{Titan's aerosol and stratospheric ice opacities
  between 18 and 500 {\ensuremath{\mu}}m: Vertical and spectral characteristics
  from Cassini CIRS}}.
\newblock \bibinfo{journal}{Icarus} \bibinfo{volume}{212},
  \bibinfo{pages}{762--778}.
\newblock \DOIprefix\doi{10.1016/j.icarus.2011.01.024}.
\bibitem[{{B{\'e}zard} et~al.(2024){B{\'e}zard}, {Greathouse} and
  {Giles}}]{Bezard2024}
\bibinfo{author}{{B{\'e}zard}, B.}, \bibinfo{author}{{Greathouse}, T.K.},
  \bibinfo{author}{{Giles}, R.}, \bibinfo{year}{2024}.
\newblock \bibinfo{title}{{The D/H ratio in Titan's acetylene from high
  spectral resolution IRTF/TEXES observations: TEXES spectra of Titan and 10
  Hygiea. Mendeley Data, v1.}} \DOIprefix\doi{10.17632/pb6f6rrypb.1}.
\bibitem[{{B{\'e}zard} et~al.(2007){B{\'e}zard}, {Nixon}, {Kleiner} and
  {Jennings}}]{Bezard2007}
\bibinfo{author}{{B{\'e}zard}, B.}, \bibinfo{author}{{Nixon}, C.A.},
  \bibinfo{author}{{Kleiner}, I.}, \bibinfo{author}{{Jennings}, D.E.},
  \bibinfo{year}{2007}.
\newblock \bibinfo{title}{{Detection of $^{13}$CH$_{3}$D on Titan}}.
\newblock \bibinfo{journal}{Icarus} \bibinfo{volume}{191},
  \bibinfo{pages}{397--400}.
\newblock \DOIprefix\doi{10.1016/j.icarus.2007.06.004}.
\bibitem[{{B{\'e}zard} and {Vinatier}(2020)}]{Bezard2020}
\bibinfo{author}{{B{\'e}zard}, B.}, \bibinfo{author}{{Vinatier}, S.},
  \bibinfo{year}{2020}.
\newblock \bibinfo{title}{{On the H$_{2}$ abundance and ortho-to-para ratio in
  Titan's troposphere}}.
\newblock \bibinfo{journal}{Icarus} \bibinfo{volume}{344},
  \bibinfo{pages}{113261}.
\newblock \DOIprefix\doi{10.1016/j.icarus.2019.03.038}.
\bibitem[{{B{\'e}zard} et~al.(2018){B{\'e}zard}, {Vinatier} and
  {Achterberg}}]{Bezard2018}
\bibinfo{author}{{B{\'e}zard}, B.}, \bibinfo{author}{{Vinatier}, S.},
  \bibinfo{author}{{Achterberg}, R.K.}, \bibinfo{year}{2018}.
\newblock \bibinfo{title}{{Seasonal radiative modeling of Titan's stratospheric
  temperatures at low latitudes}}.
\newblock \bibinfo{journal}{Icarus} \bibinfo{volume}{302},
  \bibinfo{pages}{437--450}.
\newblock \DOIprefix\doi{10.1016/j.icarus.2017.11.034}.
\bibitem[{{Borysow} and {Frommhold}(1986)}]{Borysow1986b}
\bibinfo{author}{{Borysow}, A.}, \bibinfo{author}{{Frommhold}, L.},
  \bibinfo{year}{1986}.
\newblock \bibinfo{title}{{Theoretical Collision-induced Rototranslational
  Absorption Spectra for Modeling Titan's Atmosphere: H$_2$--N$_2$ Pairs}}.
\newblock \bibinfo{journal}{Astrophys. J.} \bibinfo{volume}{303},
  \bibinfo{pages}{495}.
\newblock \DOIprefix\doi{10.1086/164096}.
\bibitem[{{Bouanich} et~al.(1998){Bouanich}, {Blanquet}, {Populaire} and
  {Walrand}}]{Bouanich1998}
\bibinfo{author}{{Bouanich}, J.P.}, \bibinfo{author}{{Blanquet}, G.},
  \bibinfo{author}{{Populaire}, J.C.}, \bibinfo{author}{{Walrand}, J.},
  \bibinfo{year}{1998}.
\newblock \bibinfo{title}{{Nitrogen Broadening of Acetylene Lines in the
  {\ensuremath{\nu}}$_{5}$ Band at Low Temperature}}.
\newblock \bibinfo{journal}{J. Molec. Spectrosc.} \bibinfo{volume}{190},
  \bibinfo{pages}{7--14}.
\newblock \DOIprefix\doi{10.1006/jmsp.1998.7559}.
\bibitem[{{Conrath} et~al.(1998){Conrath}, {Gierasch} and
  {Ustinov}}]{Conrath1998}
\bibinfo{author}{{Conrath}, B.J.}, \bibinfo{author}{{Gierasch}, P.J.},
  \bibinfo{author}{{Ustinov}, E.A.}, \bibinfo{year}{1998}.
\newblock \bibinfo{title}{{Thermal Structure and Para Hydrogen Fraction on the
  Outer Planets from Voyager IRIS Measurements}}.
\newblock \bibinfo{journal}{Icarus} \bibinfo{volume}{135},
  \bibinfo{pages}{501--517}.
\newblock \DOIprefix\doi{10.1006/icar.1998.6000}.
\bibitem[{{Courtin} et~al.(2011){Courtin}, {Swinyard}, {Moreno}, {Fulton},
  {Lellouch}, {Rengel} and {Hartogh}}]{Courtin2011}
\bibinfo{author}{{Courtin}, R.}, \bibinfo{author}{{Swinyard}, B.M.},
  \bibinfo{author}{{Moreno}, R.}, \bibinfo{author}{{Fulton}, T.},
  \bibinfo{author}{{Lellouch}, E.}, \bibinfo{author}{{Rengel}, M.},
  \bibinfo{author}{{Hartogh}, P.}, \bibinfo{year}{2011}.
\newblock \bibinfo{title}{{First results of Herschel-SPIRE observations of
  Titan}}.
\newblock \bibinfo{journal}{Astronom. Astrophys.} \bibinfo{volume}{536},
  \bibinfo{pages}{L2}.
\newblock \DOIprefix\doi{10.1051/0004-6361/201118304}.
\bibitem[{{Coustenis} et~al.(2007){Coustenis}, {Achterberg}, {Conrath},
  {Jennings}, {Marten}, {Gautier}, {Nixon}, {Flasar}, {Teanby}, {B{\'e}zard},
  {Samuelson}, {Carlson}, {Lellouch}, {Bjoraker}, {Romani}, {Taylor}, {Irwin},
  {Fouchet}, {Hubert}, {Orton}, {Kunde}, {Vinatier}, {Mondellini}, {Abbas} and
  {Courtin}}]{Coustenis2007}
\bibinfo{author}{{Coustenis}, A.}, \bibinfo{author}{{Achterberg}, R.K.},
  \bibinfo{author}{{Conrath}, B.J.}, \bibinfo{author}{{Jennings}, D.E.},
  \bibinfo{author}{{Marten}, A.}, \bibinfo{author}{{Gautier}, D.},
  \bibinfo{author}{{Nixon}, C.A.}, \bibinfo{author}{{Flasar}, F.M.},
  \bibinfo{author}{{Teanby}, N.A.}, \bibinfo{author}{{B{\'e}zard}, B.},
  \bibinfo{author}{{Samuelson}, R.E.}, \bibinfo{author}{{Carlson}, R.C.},
  \bibinfo{author}{{Lellouch}, E.}, \bibinfo{author}{{Bjoraker}, G.L.},
  \bibinfo{author}{{Romani}, P.N.}, \bibinfo{author}{{Taylor}, F.W.},
  \bibinfo{author}{{Irwin}, P.G.J.}, \bibinfo{author}{{Fouchet}, T.},
  \bibinfo{author}{{Hubert}, A.}, \bibinfo{author}{{Orton}, G.S.},
  \bibinfo{author}{{Kunde}, V.G.}, \bibinfo{author}{{Vinatier}, S.},
  \bibinfo{author}{{Mondellini}, J.}, \bibinfo{author}{{Abbas}, M.M.},
  \bibinfo{author}{{Courtin}, R.}, \bibinfo{year}{2007}.
\newblock \bibinfo{title}{{The composition of Titan's stratosphere from
  Cassini/CIRS mid-infrared spectra}}.
\newblock \bibinfo{journal}{Icarus} \bibinfo{volume}{189},
  \bibinfo{pages}{35--62}.
\newblock \DOIprefix\doi{10.1016/j.icarus.2006.12.022}.
\bibitem[{{Coustenis} et~al.(1989){Coustenis}, {B{\'e}zard} and
  {Gautier}}]{Coustenis1989}
\bibinfo{author}{{Coustenis}, A.}, \bibinfo{author}{{B{\'e}zard}, B.},
  \bibinfo{author}{{Gautier}, D.}, \bibinfo{year}{1989}.
\newblock \bibinfo{title}{{Titan's atmosphere from Voyager infrared
  observations . II. The CH$_{3}$D abundance and D/H ratio from the 900-1200
  cm$^{-1}$ spectral region}}.
\newblock \bibinfo{journal}{Icarus} \bibinfo{volume}{82},
  \bibinfo{pages}{67--80}.
\bibitem[{{Coustenis} et~al.(2020){Coustenis}, {Jennings}, {Achterberg},
  {Lavvas}, {Bampasidis}, {Nixon} and {Flasar}}]{Coustenis2020}
\bibinfo{author}{{Coustenis}, A.}, \bibinfo{author}{{Jennings}, D.E.},
  \bibinfo{author}{{Achterberg}, R.K.}, \bibinfo{author}{{Lavvas}, P.},
  \bibinfo{author}{{Bampasidis}, G.}, \bibinfo{author}{{Nixon}, C.A.},
  \bibinfo{author}{{Flasar}, F.M.}, \bibinfo{year}{2020}.
\newblock \bibinfo{title}{{Titan's neutral atmosphere seasonal variations up to
  the end of the Cassini mission}}.
\newblock \bibinfo{journal}{Icarus} \bibinfo{volume}{344},
  \bibinfo{pages}{113413}.
\newblock \DOIprefix\doi{10.1016/j.icarus.2019.113413}.
\bibitem[{{Coustenis} et~al.(2008){Coustenis}, {Jennings}, {Jolly},
  {B{\'e}nilan}, {Nixon}, {Vinatier}, {Gautier}, {Bjoraker}, {Romani},
  {Carlson} and {Flasar}}]{Coustenis2008}
\bibinfo{author}{{Coustenis}, A.}, \bibinfo{author}{{Jennings}, D.E.},
  \bibinfo{author}{{Jolly}, A.}, \bibinfo{author}{{B{\'e}nilan}, Y.},
  \bibinfo{author}{{Nixon}, C.A.}, \bibinfo{author}{{Vinatier}, S.},
  \bibinfo{author}{{Gautier}, D.}, \bibinfo{author}{{Bjoraker}, G.L.},
  \bibinfo{author}{{Romani}, P.N.}, \bibinfo{author}{{Carlson}, R.C.},
  \bibinfo{author}{{Flasar}, F.M.}, \bibinfo{year}{2008}.
\newblock \bibinfo{title}{{Detection of C$_{2}$HD and the D/H ratio on Titan}}.
\newblock \bibinfo{journal}{Icarus} \bibinfo{volume}{197},
  \bibinfo{pages}{539--548}.
\newblock \DOIprefix\doi{10.1016/j.icarus.2008.06.003}.
\bibitem[{{Coustenis} et~al.(2003){Coustenis}, {Salama}, {Schulz}, {Ott},
  {Lellouch}, {Encrenaz}, {Gautier} and {Feuchtgruber}}]{Coustenis2003}
\bibinfo{author}{{Coustenis}, A.}, \bibinfo{author}{{Salama}, A.},
  \bibinfo{author}{{Schulz}, B.}, \bibinfo{author}{{Ott}, S.},
  \bibinfo{author}{{Lellouch}, E.}, \bibinfo{author}{{Encrenaz}, T.},
  \bibinfo{author}{{Gautier}, D.}, \bibinfo{author}{{Feuchtgruber}, H.},
  \bibinfo{year}{2003}.
\newblock \bibinfo{title}{{Titan's atmosphere from ISO mid-infrared
  spectroscopy}}.
\newblock \bibinfo{journal}{Icarus} \bibinfo{volume}{161},
  \bibinfo{pages}{383--403}.
\newblock \DOIprefix\doi{10.1016/S0019-1035(02)00028-3}.
\bibitem[{{de Bergh} et~al.(2012){de Bergh}, {Courtin}, {B{\'e}zard},
  {Coustenis}, {Lellouch}, {Hirtzig}, {Rannou}, {Drossart}, {Campargue},
  {Kassi}, {Wang}, {Boudon}, {Nikitin} and {Tyuterev}}]{deBergh2012}
\bibinfo{author}{{de Bergh}, C.}, \bibinfo{author}{{Courtin}, R.},
  \bibinfo{author}{{B{\'e}zard}, B.}, \bibinfo{author}{{Coustenis}, A.},
  \bibinfo{author}{{Lellouch}, E.}, \bibinfo{author}{{Hirtzig}, M.},
  \bibinfo{author}{{Rannou}, P.}, \bibinfo{author}{{Drossart}, P.},
  \bibinfo{author}{{Campargue}, A.}, \bibinfo{author}{{Kassi}, S.},
  \bibinfo{author}{{Wang}, L.}, \bibinfo{author}{{Boudon}, V.},
  \bibinfo{author}{{Nikitin}, A.}, \bibinfo{author}{{Tyuterev}, V.},
  \bibinfo{year}{2012}.
\newblock \bibinfo{title}{{Applications of a new set of methane line parameters
  to the modeling of Titan{\textquoteright}s spectrum in the 1.58
  {\ensuremath{\mu}}m window}}.
\newblock \bibinfo{journal}{Planet. Space Sci.} \bibinfo{volume}{61},
  \bibinfo{pages}{85--98}.
\newblock \DOIprefix\doi{10.1016/j.pss.2011.05.003}.
\bibitem[{{de Bergh} et~al.(1988){de Bergh}, {Lutz}, {Owen} and
  {Chauville}}]{deBergh1988}
\bibinfo{author}{{de Bergh}, C.}, \bibinfo{author}{{Lutz}, B.L.},
  \bibinfo{author}{{Owen}, T.}, \bibinfo{author}{{Chauville}, J.},
  \bibinfo{year}{1988}.
\newblock \bibinfo{title}{{Monodeuterated Methane in the Outer Solar System.
  III. Its Abundance on Titan}}.
\newblock \bibinfo{journal}{Astrophys. J.} \bibinfo{volume}{329},
  \bibinfo{pages}{951}.
\newblock \DOIprefix\doi{10.1086/166439}.
\bibitem[{{Delahaye} et~al.(2021){Delahaye}, {Armante}, {Scott},
  {Jacquinet-Husson}, {Ch{\'e}din}, {Cr{\'e}peau}, {Crevoisier}, {Douet},
  {Perrin}, {Barbe}, {Boudon}, {Campargue}, {Coudert}, {Ebert}, {Flaud},
  {Gamache}, {Jacquemart}, {Jolly}, {Kwabia Tchana}, {Kyuberis}, {Li},
  {Lyulin}, {Manceron}, {Mikhailenko}, {Moazzen-Ahmadi}, {M{\"u}ller},
  {Naumenko}, {Nikitin}, {Perevalov}, {Richard}, {Starikova}, {Tashkun},
  {Tyuterev}, {Vander Auwera}, {Vispoel}, {Yachmenev} and
  {Yurchenko}}]{Delahaye2021}
\bibinfo{author}{{Delahaye}, T.}, \bibinfo{author}{{Armante}, R.},
  \bibinfo{author}{{Scott}, N.A.}, \bibinfo{author}{{Jacquinet-Husson}, N.},
  \bibinfo{author}{{Ch{\'e}din}, A.}, \bibinfo{author}{{Cr{\'e}peau}, L.},
  \bibinfo{author}{{Crevoisier}, C.}, \bibinfo{author}{{Douet}, V.},
  \bibinfo{author}{{Perrin}, A.}, \bibinfo{author}{{Barbe}, A.},
  \bibinfo{author}{{Boudon}, V.}, \bibinfo{author}{{Campargue}, A.},
  \bibinfo{author}{{Coudert}, L.H.}, \bibinfo{author}{{Ebert}, V.},
  \bibinfo{author}{{Flaud}, J.M.}, \bibinfo{author}{{Gamache}, R.R.},
  \bibinfo{author}{{Jacquemart}, D.}, \bibinfo{author}{{Jolly}, A.},
  \bibinfo{author}{{Kwabia Tchana}, F.}, \bibinfo{author}{{Kyuberis}, A.},
  \bibinfo{author}{{Li}, G.}, \bibinfo{author}{{Lyulin}, O.M.},
  \bibinfo{author}{{Manceron}, L.}, \bibinfo{author}{{Mikhailenko}, S.},
  \bibinfo{author}{{Moazzen-Ahmadi}, N.}, \bibinfo{author}{{M{\"u}ller},
  H.S.P.}, \bibinfo{author}{{Naumenko}, O.V.}, \bibinfo{author}{{Nikitin}, A.},
  \bibinfo{author}{{Perevalov}, V.I.}, \bibinfo{author}{{Richard}, C.},
  \bibinfo{author}{{Starikova}, E.}, \bibinfo{author}{{Tashkun}, S.A.},
  \bibinfo{author}{{Tyuterev}, V.G.}, \bibinfo{author}{{Vander Auwera}, J.},
  \bibinfo{author}{{Vispoel}, B.}, \bibinfo{author}{{Yachmenev}, A.},
  \bibinfo{author}{{Yurchenko}, S.}, \bibinfo{year}{2021}.
\newblock \bibinfo{title}{{The 2020 edition of the GEISA spectroscopic
  database}}.
\newblock \bibinfo{journal}{Journal of Molecular Spectroscopy}
  \bibinfo{volume}{380}, \bibinfo{pages}{111510}.
\newblock \DOIprefix\doi{10.1016/j.jms.2021.111510}.
\bibitem[{{Dobrijevic} et~al.(2014){Dobrijevic}, {H{\'e}brard}, {Loison} and
  {Hickson}}]{Dobrijevic2014}
\bibinfo{author}{{Dobrijevic}, M.}, \bibinfo{author}{{H{\'e}brard}, E.},
  \bibinfo{author}{{Loison}, J.C.}, \bibinfo{author}{{Hickson}, K.M.},
  \bibinfo{year}{2014}.
\newblock \bibinfo{title}{{Coupling of oxygen, nitrogen, and hydrocarbon
  species in the photochemistry of Titan's atmosphere}}.
\newblock \bibinfo{journal}{Icarus} \bibinfo{volume}{228},
  \bibinfo{pages}{324--346}.
\newblock \DOIprefix\doi{10.1016/j.icarus.2013.10.015}.
\bibitem[{{Dobrijevic} and {Loison}(2018)}]{Dobrijevic2018}
\bibinfo{author}{{Dobrijevic}, M.}, \bibinfo{author}{{Loison}, J.C.},
  \bibinfo{year}{2018}.
\newblock \bibinfo{title}{{The photochemical fractionation of nitrogen
  isotopologues in Titan's atmosphere}}.
\newblock \bibinfo{journal}{Icarus} \bibinfo{volume}{307},
  \bibinfo{pages}{371--379}.
\newblock \DOIprefix\doi{10.1016/j.icarus.2017.10.027}.
\bibitem[{{Dobrijevic} et~al.(2016){Dobrijevic}, {Loison}, {Hickson} and
  {Gronoff}}]{Dobrijevic2016}
\bibinfo{author}{{Dobrijevic}, M.}, \bibinfo{author}{{Loison}, J.C.},
  \bibinfo{author}{{Hickson}, K.M.}, \bibinfo{author}{{Gronoff}, G.},
  \bibinfo{year}{2016}.
\newblock \bibinfo{title}{{1D-coupled photochemical model of neutrals, cations
  and anions in the atmosphere of Titan}}.
\newblock \bibinfo{journal}{Icarus} \bibinfo{volume}{268},
  \bibinfo{pages}{313--339}.
\newblock \DOIprefix\doi{10.1016/j.icarus.2015.12.045}.
\bibitem[{{Finenko} et~al.(2022){Finenko}, {B{\'e}zard}, {Gordon}, {Chistikov},
  {Lokshtanov}, {Petrov} and {Vigasin}}]{Finenko2022}
\bibinfo{author}{{Finenko}, A.A.}, \bibinfo{author}{{B{\'e}zard}, B.},
  \bibinfo{author}{{Gordon}, I.E.}, \bibinfo{author}{{Chistikov}, D.N.},
  \bibinfo{author}{{Lokshtanov}, S.E.}, \bibinfo{author}{{Petrov}, S.V.},
  \bibinfo{author}{{Vigasin}, A.A.}, \bibinfo{year}{2022}.
\newblock \bibinfo{title}{{Trajectory-based Simulation of Far-infrared
  Collision-induced Absorption Profiles of CH$_{4}$-N$_{2}$ for Modeling
  Titan's Atmosphere}}.
\newblock \bibinfo{journal}{Astrophys. J. Suppl. S} \bibinfo{volume}{258},
  \bibinfo{pages}{33}.
\newblock \DOIprefix\doi{10.3847/1538-4365/ac36d3}.
\bibitem[{{Flasar} et~al.(2004){Flasar}, {Kunde}, {Abbas}, {Achterberg}, {Ade},
  {Barucci}, {B{\'e}zard}, {Bjoraker}, {Brasunas}, {Calcutt}, {Carlson},
  {C{\'e}sarsky}, {Conrath}, {Coradini}, {Courtin}, {Coustenis}, {Edberg},
  {Edgington}, {Ferrari}, {Fouchet}, {Gautier}, {Gierasch}, {Grossman},
  {Irwin}, {Jennings}, {Lellouch}, {Mamoutkine}, {Marten}, {Meyer}, {Nixon},
  {Orton}, {Owen}, {Pearl}, {Prang{\'e}}, {Raulin}, {Read}, {Romani},
  {Samuelson}, {Segura}, {Showalter}, {Simon-Miller}, {Smith}, {Spencer},
  {Spilker} and {Taylor}}]{Flasar2004}
\bibinfo{author}{{Flasar}, F.M.}, \bibinfo{author}{{Kunde}, V.G.},
  \bibinfo{author}{{Abbas}, M.M.}, \bibinfo{author}{{Achterberg}, R.K.},
  \bibinfo{author}{{Ade}, P.}, \bibinfo{author}{{Barucci}, A.},
  \bibinfo{author}{{B{\'e}zard}, B.}, \bibinfo{author}{{Bjoraker}, G.L.},
  \bibinfo{author}{{Brasunas}, J.C.}, \bibinfo{author}{{Calcutt}, S.},
  \bibinfo{author}{{Carlson}, R.}, \bibinfo{author}{{C{\'e}sarsky}, C.J.},
  \bibinfo{author}{{Conrath}, B.J.}, \bibinfo{author}{{Coradini}, A.},
  \bibinfo{author}{{Courtin}, R.}, \bibinfo{author}{{Coustenis}, A.},
  \bibinfo{author}{{Edberg}, S.}, \bibinfo{author}{{Edgington}, S.},
  \bibinfo{author}{{Ferrari}, C.}, \bibinfo{author}{{Fouchet}, T.},
  \bibinfo{author}{{Gautier}, D.}, \bibinfo{author}{{Gierasch}, P.J.},
  \bibinfo{author}{{Grossman}, K.}, \bibinfo{author}{{Irwin}, P.},
  \bibinfo{author}{{Jennings}, D.E.}, \bibinfo{author}{{Lellouch}, E.},
  \bibinfo{author}{{Mamoutkine}, A.A.}, \bibinfo{author}{{Marten}, A.},
  \bibinfo{author}{{Meyer}, J.P.}, \bibinfo{author}{{Nixon}, C.A.},
  \bibinfo{author}{{Orton}, G.S.}, \bibinfo{author}{{Owen}, T.C.},
  \bibinfo{author}{{Pearl}, J.C.}, \bibinfo{author}{{Prang{\'e}}, R.},
  \bibinfo{author}{{Raulin}, F.}, \bibinfo{author}{{Read}, P.L.},
  \bibinfo{author}{{Romani}, P.N.}, \bibinfo{author}{{Samuelson}, R.E.},
  \bibinfo{author}{{Segura}, M.E.}, \bibinfo{author}{{Showalter}, M.R.},
  \bibinfo{author}{{Simon-Miller}, A.A.}, \bibinfo{author}{{Smith}, M.D.},
  \bibinfo{author}{{Spencer}, J.R.}, \bibinfo{author}{{Spilker}, L.J.},
  \bibinfo{author}{{Taylor}, F.W.}, \bibinfo{year}{2004}.
\newblock \bibinfo{title}{{Exploring The Saturn System In The Thermal Infrared:
  The Composite Infrared Spectrometer}}.
\newblock \bibinfo{journal}{Space Sci. Rev.} \bibinfo{volume}{115},
  \bibinfo{pages}{169--297}.
\newblock \DOIprefix\doi{10.1007/s11214-004-1454-9}.
\bibitem[{{Fulchignoni} et~al.(2005){Fulchignoni}, {Ferri}, {Angrilli}, {Ball},
  {Bar-Nun}, {Barucci}, {Bettanini}, {Bianchini}, {Borucki}, {Colombatti},
  {Coradini}, {Coustenis}, {Debei}, {Falkner}, {Fanti}, {Flamini}, {Gaborit},
  {Grard}, {Hamelin}, {Harri}, {Hathi}, {Jernej}, {Leese}, {Lehto}, {Lion
  Stoppato}, {L{\'o}pez-Moreno}, {M{\"a}kinen}, {McDonnell}, {McKay},
  {Molina-Cuberos}, {Neubauer}, {Pirronello}, {Rodrigo}, {Saggin},
  {Schwingenschuh}, {Seiff}, {Sim{\~o}es}, {Svedhem}, {Tokano}, {Towner},
  {Trautner}, {Withers} and {Zarnecki}}]{Fulchignoni2005}
\bibinfo{author}{{Fulchignoni}, M.}, \bibinfo{author}{{Ferri}, F.},
  \bibinfo{author}{{Angrilli}, F.}, \bibinfo{author}{{Ball}, A.J.},
  \bibinfo{author}{{Bar-Nun}, A.}, \bibinfo{author}{{Barucci}, M.A.},
  \bibinfo{author}{{Bettanini}, C.}, \bibinfo{author}{{Bianchini}, G.},
  \bibinfo{author}{{Borucki}, W.}, \bibinfo{author}{{Colombatti}, G.},
  \bibinfo{author}{{Coradini}, M.}, \bibinfo{author}{{Coustenis}, A.},
  \bibinfo{author}{{Debei}, S.}, \bibinfo{author}{{Falkner}, P.},
  \bibinfo{author}{{Fanti}, G.}, \bibinfo{author}{{Flamini}, E.},
  \bibinfo{author}{{Gaborit}, V.}, \bibinfo{author}{{Grard}, R.},
  \bibinfo{author}{{Hamelin}, M.}, \bibinfo{author}{{Harri}, A.M.},
  \bibinfo{author}{{Hathi}, B.}, \bibinfo{author}{{Jernej}, I.},
  \bibinfo{author}{{Leese}, M.R.}, \bibinfo{author}{{Lehto}, A.},
  \bibinfo{author}{{Lion Stoppato}, P.F.}, \bibinfo{author}{{L{\'o}pez-Moreno},
  J.J.}, \bibinfo{author}{{M{\"a}kinen}, T.}, \bibinfo{author}{{McDonnell},
  J.A.M.}, \bibinfo{author}{{McKay}, C.P.}, \bibinfo{author}{{Molina-Cuberos},
  G.}, \bibinfo{author}{{Neubauer}, F.M.}, \bibinfo{author}{{Pirronello}, V.},
  \bibinfo{author}{{Rodrigo}, R.}, \bibinfo{author}{{Saggin}, B.},
  \bibinfo{author}{{Schwingenschuh}, K.}, \bibinfo{author}{{Seiff}, A.},
  \bibinfo{author}{{Sim{\~o}es}, F.}, \bibinfo{author}{{Svedhem}, H.},
  \bibinfo{author}{{Tokano}, T.}, \bibinfo{author}{{Towner}, M.C.},
  \bibinfo{author}{{Trautner}, R.}, \bibinfo{author}{{Withers}, P.},
  \bibinfo{author}{{Zarnecki}, J.C.}, \bibinfo{year}{2005}.
\newblock \bibinfo{title}{{In situ measurements of the physical characteristics
  of Titan's environment}}.
\newblock \bibinfo{journal}{Nature} \bibinfo{volume}{438},
  \bibinfo{pages}{785--791}.
\newblock \DOIprefix\doi{10.1038/nature04314}.
\bibitem[{{Gurwell}(2004)}]{Gurwell2004}
\bibinfo{author}{{Gurwell}, M.A.}, \bibinfo{year}{2004}.
\newblock \bibinfo{title}{{Submillimeter Observations of Titan: Global Measures
  of Stratospheric Temperature, CO, HCN, HC$_{3}$N, and the Isotopic Ratios
  $^{12}$C/$^{13}$C and $^{14}$N/$^{15}$N}}.
\newblock \bibinfo{journal}{Astrophys. J. Lett.} \bibinfo{volume}{616},
  \bibinfo{pages}{L7--L10}.
\newblock \DOIprefix\doi{10.1086/423954}.
\bibitem[{{Jennings} et~al.(2017){Jennings}, {Flasar}, {Kunde}, {Nixon},
  {Segura}, {Romani}, {Gorius}, {Albright}, {Brasunas}, {Carlson},
  {Mamoutkine}, {Guandique}, {Kaelberer}, {Aslam}, {Achterberg}, {Bjoraker},
  {Anderson}, {Cottini}, {Pearl}, {Smith}, {Hesman}, {Barney}, {Calcutt},
  {Vellacott}, {Spilker}, {Edgington}, {Brooks}, {Ade}, {Schinder},
  {Coustenis}, {Courtin}, {Michel}, {Fettig}, {Pilorz} and
  {Ferrari}}]{Jennings2017}
\bibinfo{author}{{Jennings}, D.E.}, \bibinfo{author}{{Flasar}, F.M.},
  \bibinfo{author}{{Kunde}, V.G.}, \bibinfo{author}{{Nixon}, C.A.},
  \bibinfo{author}{{Segura}, M.E.}, \bibinfo{author}{{Romani}, P.N.},
  \bibinfo{author}{{Gorius}, N.}, \bibinfo{author}{{Albright}, S.},
  \bibinfo{author}{{Brasunas}, J.C.}, \bibinfo{author}{{Carlson}, R.C.},
  \bibinfo{author}{{Mamoutkine}, A.A.}, \bibinfo{author}{{Guandique}, E.},
  \bibinfo{author}{{Kaelberer}, M.S.}, \bibinfo{author}{{Aslam}, S.},
  \bibinfo{author}{{Achterberg}, R.K.}, \bibinfo{author}{{Bjoraker}, G.L.},
  \bibinfo{author}{{Anderson}, C.M.}, \bibinfo{author}{{Cottini}, V.},
  \bibinfo{author}{{Pearl}, J.C.}, \bibinfo{author}{{Smith}, M.D.},
  \bibinfo{author}{{Hesman}, B.E.}, \bibinfo{author}{{Barney}, R.D.},
  \bibinfo{author}{{Calcutt}, S.}, \bibinfo{author}{{Vellacott}, T.J.},
  \bibinfo{author}{{Spilker}, L.J.}, \bibinfo{author}{{Edgington}, S.G.},
  \bibinfo{author}{{Brooks}, S.M.}, \bibinfo{author}{{Ade}, P.},
  \bibinfo{author}{{Schinder}, P.J.}, \bibinfo{author}{{Coustenis}, A.},
  \bibinfo{author}{{Courtin}, R.}, \bibinfo{author}{{Michel}, G.},
  \bibinfo{author}{{Fettig}, R.}, \bibinfo{author}{{Pilorz}, S.},
  \bibinfo{author}{{Ferrari}, C.}, \bibinfo{year}{2017}.
\newblock \bibinfo{title}{{Composite infrared spectrometer (CIRS) on Cassini}}.
\newblock \bibinfo{journal}{Appl. Optics} \bibinfo{volume}{56},
  \bibinfo{pages}{5274}.
\newblock \DOIprefix\doi{10.1364/ao.56.005274}.
\bibitem[{{Jolly} et~al.(2008){Jolly}, {Benilan}, {Can{\'e}}, {Fusina},
  {Tamassia}, {Fayt}, {Robert} and {Herman}}]{Jolly2008}
\bibinfo{author}{{Jolly}, A.}, \bibinfo{author}{{Benilan}, Y.},
  \bibinfo{author}{{Can{\'e}}, E.}, \bibinfo{author}{{Fusina}, L.},
  \bibinfo{author}{{Tamassia}, F.}, \bibinfo{author}{{Fayt}, A.},
  \bibinfo{author}{{Robert}, S.}, \bibinfo{author}{{Herman}, M.},
  \bibinfo{year}{2008}.
\newblock \bibinfo{title}{{Measured integrated band intensities and simulated
  line-by-line spectra for $^{12}$C$_{2}$HD between 25 and 2.5
  {\ensuremath{\mu}}m, and new global vibration rotation parameters for the
  bending vibrations}}.
\newblock \bibinfo{journal}{J. Quant. Spectrosc. RA} \bibinfo{volume}{109},
  \bibinfo{pages}{2846--2856}.
\newblock \DOIprefix\doi{10.1016/j.jqsrt.2008.08.004}.
\bibitem[{{Karman} et~al.(2015){Karman}, {Miliordos}, {Hunt}, {Groenenboom} and
  {van der Avoird}}]{Karman2015}
\bibinfo{author}{{Karman}, T.}, \bibinfo{author}{{Miliordos}, E.},
  \bibinfo{author}{{Hunt}, K.L.C.}, \bibinfo{author}{{Groenenboom}, G.C.},
  \bibinfo{author}{{van der Avoird}, A.}, \bibinfo{year}{2015}.
\newblock \bibinfo{title}{{Quantum mechanical calculation of the
  collision-induced absorption spectra of N$_{2}$-N$_{2}$ with anisotropic
  interactions}}.
\newblock \bibinfo{journal}{J. Chem. Phys.} \bibinfo{volume}{142},
  \bibinfo{pages}{084306}.
\newblock \DOIprefix\doi{10.1063/1.4907917}.
\bibitem[{{Krasnopolsky}(2014)}]{Krasnopolsky2014}
\bibinfo{author}{{Krasnopolsky}, V.A.}, \bibinfo{year}{2014}.
\newblock \bibinfo{title}{{Chemical composition of Titan{\textquoteright}s
  atmosphere and ionosphere: Observations and the photochemical model}}.
\newblock \bibinfo{journal}{Icarus} \bibinfo{volume}{236},
  \bibinfo{pages}{83--91}.
\newblock \DOIprefix\doi{10.1016/j.icarus.2014.03.041}.
\bibitem[{{Lacy} et~al.(2002){Lacy}, {Richter}, {Greathouse}, {Jaffe} and
  {Zhu}}]{Lacy2002}
\bibinfo{author}{{Lacy}, J.H.}, \bibinfo{author}{{Richter}, M.J.},
  \bibinfo{author}{{Greathouse}, T.K.}, \bibinfo{author}{{Jaffe}, D.T.},
  \bibinfo{author}{{Zhu}, Q.}, \bibinfo{year}{2002}.
\newblock \bibinfo{title}{{TEXES: A Sensitive High-Resolution Grating
  Spectrograph for the Mid-Infrared}}.
\newblock \bibinfo{journal}{Publ. Astron. Soc. Pac.} \bibinfo{volume}{114},
  \bibinfo{pages}{153--168}.
\newblock \DOIprefix\doi{10.1086/338730}.
\bibitem[{{Lara} et~al.(2014){Lara}, {Lellouch}, {Gonz{\'a}lez}, {Moreno} and
  {Rengel}}]{Lara2014}
\bibinfo{author}{{Lara}, L.M.}, \bibinfo{author}{{Lellouch}, E.},
  \bibinfo{author}{{Gonz{\'a}lez}, M.}, \bibinfo{author}{{Moreno}, R.},
  \bibinfo{author}{{Rengel}, M.}, \bibinfo{year}{2014}.
\newblock \bibinfo{title}{{A time-dependent photochemical model for Titan's
  atmosphere and the origin of H$_{2}$O}}.
\newblock \bibinfo{journal}{Astronom. Astrophys.} \bibinfo{volume}{566},
  \bibinfo{pages}{A143}.
\newblock \DOIprefix\doi{10.1051/0004-6361/201323085}.
\bibitem[{Lebofsky and Spencer(1989)}]{Lebofsky1989}
\bibinfo{author}{Lebofsky, L.}, \bibinfo{author}{Spencer, J.},
  \bibinfo{year}{1989}.
\newblock \bibinfo{title}{Radiometry and thermal modeling of asteroids}, in:
  \bibinfo{editor}{Binzel, R.}, \bibinfo{editor}{Gehrels, T.},
  \bibinfo{editor}{Matthews, M.} (Eds.), \bibinfo{booktitle}{Asteroids II}.
  \bibinfo{publisher}{Univ. of Arizona Press, Tucson}, pp.
  \bibinfo{pages}{128--147}.
\bibitem[{{Lellouch} et~al.(2014){Lellouch}, {B{\'e}zard}, {Flasar},
  {Vinatier}, {Achterberg}, {Nixon}, {Bjoraker} and {Gorius}}]{Lellouch2014}
\bibinfo{author}{{Lellouch}, E.}, \bibinfo{author}{{B{\'e}zard}, B.},
  \bibinfo{author}{{Flasar}, F.M.}, \bibinfo{author}{{Vinatier}, S.},
  \bibinfo{author}{{Achterberg}, R.}, \bibinfo{author}{{Nixon}, C.A.},
  \bibinfo{author}{{Bjoraker}, G.L.}, \bibinfo{author}{{Gorius}, N.},
  \bibinfo{year}{2014}.
\newblock \bibinfo{title}{{The distribution of methane in
  Titan{\textquoteright}s stratosphere from Cassini/CIRS observations}}.
\newblock \bibinfo{journal}{Icarus} \bibinfo{volume}{231},
  \bibinfo{pages}{323--337}.
\newblock \DOIprefix\doi{10.1016/j.icarus.2013.12.016}.
\bibitem[{{Lim} et~al.(2005){Lim}, {McConnochie}, {Bell} and
  {Hayward}}]{Lim2005}
\bibinfo{author}{{Lim}, L.F.}, \bibinfo{author}{{McConnochie}, T.H.},
  \bibinfo{author}{{Bell}, J.F.}, \bibinfo{author}{{Hayward}, T.L.},
  \bibinfo{year}{2005}.
\newblock \bibinfo{title}{{Thermal infrared (8--13 {\ensuremath{\mu}}m) spectra
  of 29 asteroids: the Cornell Mid-Infrared Asteroid Spectroscopy (MIDAS)
  Survey}}.
\newblock \bibinfo{journal}{Icarus} \bibinfo{volume}{173},
  \bibinfo{pages}{385--408}.
\newblock \DOIprefix\doi{10.1016/j.icarus.2004.08.005}.
\bibitem[{{Lombardo} et~al.(2019){Lombardo}, {Nixon}, {Greathouse},
  {B{\'e}zard}, {Jolly}, {Vinatier}, {Teanby}, {Richter}, {Irwin}, {Coustenis}
  and {Flasar}}]{Lombardo2019}
\bibinfo{author}{{Lombardo}, N.A.}, \bibinfo{author}{{Nixon}, C.A.},
  \bibinfo{author}{{Greathouse}, T.K.}, \bibinfo{author}{{B{\'e}zard}, B.},
  \bibinfo{author}{{Jolly}, A.}, \bibinfo{author}{{Vinatier}, S.},
  \bibinfo{author}{{Teanby}, N.A.}, \bibinfo{author}{{Richter}, M.J.},
  \bibinfo{author}{{Irwin}, P.J.G.}, \bibinfo{author}{{Coustenis}, A.},
  \bibinfo{author}{{Flasar}, F.M.}, \bibinfo{year}{2019}.
\newblock \bibinfo{title}{{Detection of Propadiene on Titan}}.
\newblock \bibinfo{journal}{Astrophys. J. Lett.} \bibinfo{volume}{881},
  \bibinfo{pages}{L33}.
\newblock \DOIprefix\doi{10.3847/2041-8213/ab3860}.
\bibitem[{{Lunine} et~al.(1999){Lunine}, {Yung} and {Lorenz}}]{Lunine1999}
\bibinfo{author}{{Lunine}, J.I.}, \bibinfo{author}{{Yung}, Y.L.},
  \bibinfo{author}{{Lorenz}, R.D.}, \bibinfo{year}{1999}.
\newblock \bibinfo{title}{{On the volatile inventory of Titan from isotopic
  abundances in nitrogen and methane}}.
\newblock \bibinfo{journal}{Planet. Space Sci.} \bibinfo{volume}{47},
  \bibinfo{pages}{1291--1303}.
\newblock \DOIprefix\doi{10.1016/S0032-0633(99)00052-5}.
\bibitem[{{Marten} et~al.(2002){Marten}, {Hidayat}, {Biraud} and
  {Moreno}}]{Marten2002}
\bibinfo{author}{{Marten}, A.}, \bibinfo{author}{{Hidayat}, T.},
  \bibinfo{author}{{Biraud}, Y.}, \bibinfo{author}{{Moreno}, R.},
  \bibinfo{year}{2002}.
\newblock \bibinfo{title}{{New Millimeter Heterodyne Observations of Titan:
  Vertical Distributions of Nitriles HCN, HC $_{3}$N, CH $_{3}$CN, and the
  Isotopic Ratio $^{15}$N/ $^{14}$N in Its Atmosphere}}.
\newblock \bibinfo{journal}{Icarus} \bibinfo{volume}{158},
  \bibinfo{pages}{532--544}.
\newblock \DOIprefix\doi{10.1006/icar.2002.6897}.
\bibitem[{{Math{\'e}} et~al.(2020){Math{\'e}}, {Vinatier}, {B{\'e}zard},
  {Lebonnois}, {Gorius}, {Jennings}, {Mamoutkine}, {Guandique} and {Vatant
  d'Ollone}}]{Mathe2020}
\bibinfo{author}{{Math{\'e}}, C.}, \bibinfo{author}{{Vinatier}, S.},
  \bibinfo{author}{{B{\'e}zard}, B.}, \bibinfo{author}{{Lebonnois}, S.},
  \bibinfo{author}{{Gorius}, N.}, \bibinfo{author}{{Jennings}, D.E.},
  \bibinfo{author}{{Mamoutkine}, A.}, \bibinfo{author}{{Guandique}, E.},
  \bibinfo{author}{{Vatant d'Ollone}, J.}, \bibinfo{year}{2020}.
\newblock \bibinfo{title}{{Seasonal changes in the middle atmosphere of Titan
  from Cassini/CIRS observations: Temperature and trace species abundance
  profiles from 2004 to 2017}}.
\newblock \bibinfo{journal}{Icarus} \bibinfo{volume}{344},
  \bibinfo{pages}{113547}.
\newblock \DOIprefix\doi{10.1016/j.icarus.2019.113547}.
\bibitem[{{Moreno} et~al.(2011){Moreno}, {Lellouch}, {Lara}, {Courtin},
  {Bockel{\'e}e-Morvan}, {Hartogh}, {Rengel}, {Biver}, {Banaszkiewicz} and
  {Gonz{\'a}lez}}]{Moreno2011}
\bibinfo{author}{{Moreno}, R.}, \bibinfo{author}{{Lellouch}, E.},
  \bibinfo{author}{{Lara}, L.M.}, \bibinfo{author}{{Courtin}, R.},
  \bibinfo{author}{{Bockel{\'e}e-Morvan}, D.}, \bibinfo{author}{{Hartogh}, P.},
  \bibinfo{author}{{Rengel}, M.}, \bibinfo{author}{{Biver}, N.},
  \bibinfo{author}{{Banaszkiewicz}, M.}, \bibinfo{author}{{Gonz{\'a}lez}, A.},
  \bibinfo{year}{2011}.
\newblock \bibinfo{title}{{First detection of hydrogen isocyanide (HNC) in
  Titan's atmosphere}}.
\newblock \bibinfo{journal}{Astronom. Astrophys.} \bibinfo{volume}{536},
  \bibinfo{pages}{L12}.
\newblock \DOIprefix\doi{10.1051/0004-6361/201118189}.
\bibitem[{{Moreno} et~al.(2012){Moreno}, {Lellouch}, {Lara}, {Feuchtgruber},
  {Rengel}, {Hartogh} and {Courtin}}]{Moreno2012}
\bibinfo{author}{{Moreno}, R.}, \bibinfo{author}{{Lellouch}, E.},
  \bibinfo{author}{{Lara}, L.M.}, \bibinfo{author}{{Feuchtgruber}, H.},
  \bibinfo{author}{{Rengel}, M.}, \bibinfo{author}{{Hartogh}, P.},
  \bibinfo{author}{{Courtin}, R.}, \bibinfo{year}{2012}.
\newblock \bibinfo{title}{{The abundance, vertical distribution and origin of
  H$_{2}$O in Titan{\textquoteright}s atmosphere: Herschel observations and
  photochemical modelling}}.
\newblock \bibinfo{journal}{Icarus} \bibinfo{volume}{221},
  \bibinfo{pages}{753--767}.
\newblock \DOIprefix\doi{10.1016/j.icarus.2012.09.006}.
\bibitem[{{Niemann} et~al.(2010){Niemann}, {Atreya}, {Demick}, {Gautier},
  {Haberman}, {Harpold}, {Kasprzak}, {Lunine}, {Owen} and
  {Raulin}}]{Niemann2010}
\bibinfo{author}{{Niemann}, H.B.}, \bibinfo{author}{{Atreya}, S.K.},
  \bibinfo{author}{{Demick}, J.E.}, \bibinfo{author}{{Gautier}, D.},
  \bibinfo{author}{{Haberman}, J.A.}, \bibinfo{author}{{Harpold}, D.N.},
  \bibinfo{author}{{Kasprzak}, W.T.}, \bibinfo{author}{{Lunine}, J.I.},
  \bibinfo{author}{{Owen}, T.C.}, \bibinfo{author}{{Raulin}, F.},
  \bibinfo{year}{2010}.
\newblock \bibinfo{title}{{Composition of Titan's lower atmosphere and simple
  surface volatiles as measured by the Cassini-Huygens probe gas chromatograph
  mass spectrometer experiment}}.
\newblock \bibinfo{journal}{J. Geophys. Res.} \bibinfo{volume}{115},
  \bibinfo{pages}{E12006}.
\newblock \DOIprefix\doi{10.1029/2010JE003659}.
\bibitem[{{Nixon} et~al.(2019){Nixon}, {Ansty}, {Lombardo}, {Bjoraker},
  {Achterberg}, {Annex}, {Rice}, {Romani}, {Jennings}, {Samuelson}, {Anderson},
  {Coustenis}, {B{\'e}zard}, {Vinatier}, {Lellouch}, {Courtin}, {Teanby},
  {Cottini} and {Flasar}}]{Nixon2019}
\bibinfo{author}{{Nixon}, C.A.}, \bibinfo{author}{{Ansty}, T.M.},
  \bibinfo{author}{{Lombardo}, N.A.}, \bibinfo{author}{{Bjoraker}, G.L.},
  \bibinfo{author}{{Achterberg}, R.K.}, \bibinfo{author}{{Annex}, A.M.},
  \bibinfo{author}{{Rice}, M.}, \bibinfo{author}{{Romani}, P.N.},
  \bibinfo{author}{{Jennings}, D.E.}, \bibinfo{author}{{Samuelson}, R.E.},
  \bibinfo{author}{{Anderson}, C.M.}, \bibinfo{author}{{Coustenis}, A.},
  \bibinfo{author}{{B{\'e}zard}, B.}, \bibinfo{author}{{Vinatier}, S.},
  \bibinfo{author}{{Lellouch}, E.}, \bibinfo{author}{{Courtin}, R.},
  \bibinfo{author}{{Teanby}, N.A.}, \bibinfo{author}{{Cottini}, V.},
  \bibinfo{author}{{Flasar}, F.M.}, \bibinfo{year}{2019}.
\newblock \bibinfo{title}{{Cassini Composite Infrared Spectrometer (CIRS)
  Observations of Titan 2004-2017}}.
\newblock \bibinfo{journal}{Astrophys. J. Suppl. S} \bibinfo{volume}{244},
  \bibinfo{pages}{14}.
\newblock \DOIprefix\doi{10.3847/1538-4365/ab3799}.
\bibitem[{{Nixon} et~al.(2012){Nixon}, {Temelso}, {Vinatier}, {Teanby},
  {B{\'e}zard}, {Achterberg}, {Mandt}, {Sherrill}, {Irwin}, {Jennings},
  {Romani}, {Coustenis} and {Flasar}}]{Nixon2012}
\bibinfo{author}{{Nixon}, C.A.}, \bibinfo{author}{{Temelso}, B.},
  \bibinfo{author}{{Vinatier}, S.}, \bibinfo{author}{{Teanby}, N.A.},
  \bibinfo{author}{{B{\'e}zard}, B.}, \bibinfo{author}{{Achterberg}, R.K.},
  \bibinfo{author}{{Mandt}, K.E.}, \bibinfo{author}{{Sherrill}, C.D.},
  \bibinfo{author}{{Irwin}, P.G.J.}, \bibinfo{author}{{Jennings}, D.E.},
  \bibinfo{author}{{Romani}, P.N.}, \bibinfo{author}{{Coustenis}, A.},
  \bibinfo{author}{{Flasar}, F.M.}, \bibinfo{year}{2012}.
\newblock \bibinfo{title}{{Isotopic Ratios in Titan's Methane: Measurements and
  Modeling}}.
\newblock \bibinfo{journal}{Astrophys. J.} \bibinfo{volume}{749},
  \bibinfo{pages}{159}.
\newblock \DOIprefix\doi{10.1088/0004-637X/749/2/159}.
\bibitem[{{Nixon} et~al.(2020){Nixon}, {Thelen}, {Cordiner}, {Kisiel},
  {Charnley}, {Molter}, {Serigano}, {Irwin}, {Teanby} and {Kuan}}]{Nixon2020}
\bibinfo{author}{{Nixon}, C.A.}, \bibinfo{author}{{Thelen}, A.E.},
  \bibinfo{author}{{Cordiner}, M.A.}, \bibinfo{author}{{Kisiel}, Z.},
  \bibinfo{author}{{Charnley}, S.B.}, \bibinfo{author}{{Molter}, E.M.},
  \bibinfo{author}{{Serigano}, J.}, \bibinfo{author}{{Irwin}, P.G.J.},
  \bibinfo{author}{{Teanby}, N.A.}, \bibinfo{author}{{Kuan}, Y.J.},
  \bibinfo{year}{2020}.
\newblock \bibinfo{title}{{Detection of Cyclopropenylidene on Titan with
  ALMA}}.
\newblock \bibinfo{journal}{Astron. J.} \bibinfo{volume}{160},
  \bibinfo{pages}{205}.
\newblock \DOIprefix\doi{10.3847/1538-3881/abb679}.
\bibitem[{{Owen} et~al.(1986){Owen}, {Lutz} and {de Bergh}}]{Owen1986}
\bibinfo{author}{{Owen}, T.}, \bibinfo{author}{{Lutz}, B.L.},
  \bibinfo{author}{{de Bergh}, C.}, \bibinfo{year}{1986}.
\newblock \bibinfo{title}{{Deuterium in the outer Solar System: evidence for
  two distinct reservoirs}}.
\newblock \bibinfo{journal}{Nature} \bibinfo{volume}{320},
  \bibinfo{pages}{244--246}.
\newblock \DOIprefix\doi{10.1038/320244a0}.
\bibitem[{{Penteado} et~al.(2005){Penteado}, {Griffith}, {Greathouse} and {de
  Bergh}}]{Penteado2005}
\bibinfo{author}{{Penteado}, P.F.}, \bibinfo{author}{{Griffith}, C.A.},
  \bibinfo{author}{{Greathouse}, T.K.}, \bibinfo{author}{{de Bergh}, C.},
  \bibinfo{year}{2005}.
\newblock \bibinfo{title}{{Measurements of CH$_{3}$D and CH$_{4}$ in Titan from
  Infrared Spectroscopy}}.
\newblock \bibinfo{journal}{Astrophys. J. Lett.} \bibinfo{volume}{629},
  \bibinfo{pages}{L53--L56}.
\newblock \DOIprefix\doi{10.1086/444353}.
\bibitem[{{Pinto} et~al.(1986){Pinto}, {Lunine}, {Kim} and {Yung}}]{Pinto1986}
\bibinfo{author}{{Pinto}, J.P.}, \bibinfo{author}{{Lunine}, J.I.},
  \bibinfo{author}{{Kim}, S.J.}, \bibinfo{author}{{Yung}, Y.L.},
  \bibinfo{year}{1986}.
\newblock \bibinfo{title}{{D to H ratio and the origin and evolution of Titan's
  atmosphere}}.
\newblock \bibinfo{journal}{Nature} \bibinfo{volume}{319},
  \bibinfo{pages}{388--390}.
\newblock \DOIprefix\doi{10.1038/319388a0}.
\bibitem[{{Rey} et~al.(2018){Rey}, {Nikitin}, {B{\'e}zard}, {Rannou},
  {Coustenis} and {Tyuterev}}]{Rey2018}
\bibinfo{author}{{Rey}, M.}, \bibinfo{author}{{Nikitin}, A.V.},
  \bibinfo{author}{{B{\'e}zard}, B.}, \bibinfo{author}{{Rannou}, P.},
  \bibinfo{author}{{Coustenis}, A.}, \bibinfo{author}{{Tyuterev}, V.G.},
  \bibinfo{year}{2018}.
\newblock \bibinfo{title}{{New accurate theoretical line lists of
  $^{12}$CH$_{4}$ and $^{13}$CH$_{4}$ in the 0-13400 cm$^{-1}$ range:
  Application to the modeling of methane absorption in Titan's atmosphere}}.
\newblock \bibinfo{journal}{Icarus} \bibinfo{volume}{303},
  \bibinfo{pages}{114--130}.
\newblock \DOIprefix\doi{10.1016/j.icarus.2017.12.045}.
\bibitem[{{Thelen} et~al.(2020){Thelen}, {Cordiner}, {Nixon}, {Vuitton},
  {Kisiel}, {Charnley}, {Palmer}, {Teanby} and {Irwin}}]{Thelen2020}
\bibinfo{author}{{Thelen}, A.E.}, \bibinfo{author}{{Cordiner}, M.A.},
  \bibinfo{author}{{Nixon}, C.A.}, \bibinfo{author}{{Vuitton}, V.},
  \bibinfo{author}{{Kisiel}, Z.}, \bibinfo{author}{{Charnley}, S.B.},
  \bibinfo{author}{{Palmer}, M.Y.}, \bibinfo{author}{{Teanby}, N.A.},
  \bibinfo{author}{{Irwin}, P.G.J.}, \bibinfo{year}{2020}.
\newblock \bibinfo{title}{{Detection of CH$_{3}$C$_{3}$N in Titan's
  Atmosphere}}.
\newblock \bibinfo{journal}{Astrophys. J. Lett.} \bibinfo{volume}{903},
  \bibinfo{pages}{L22}.
\newblock \DOIprefix\doi{10.3847/2041-8213/abc1e1}.
\bibitem[{{Tribbett} et~al.(2021){Tribbett}, {Robinson} and
  {Koskinen}}]{Tribett2021}
\bibinfo{author}{{Tribbett}, P.D.}, \bibinfo{author}{{Robinson}, T.D.},
  \bibinfo{author}{{Koskinen}, T.T.}, \bibinfo{year}{2021}.
\newblock \bibinfo{title}{{Titan in Transit: Ultraviolet Stellar Occultation
  Observations Reveal a Complex Atmospheric Structure}}.
\newblock \bibinfo{journal}{Planet. Sci. J.} \bibinfo{volume}{2},
  \bibinfo{pages}{109}.
\newblock \DOIprefix\doi{10.3847/PSJ/abf92d}.
\bibitem[{{Vernazza} et~al.(2021){Vernazza}, {Ferrais}, {Jorda}, {Hanu{\v{s}}},
  {Carry}, {Marsset}, {Bro{\v{z}}}, {Fetick}, {Viikinkoski}, {Marchis},
  {Vachier}, {Drouard}, {Fusco}, {Birlan}, {Podlewska-Gaca}, {Rambaux},
  {Neveu}, {Bartczak}, {Dudzi{\'n}ski}, {Jehin}, {Beck}, {Berthier},
  {Castillo-Rogez}, {Cipriani}, {Colas}, {Dumas}, {{\v{D}}urech}, {Grice},
  {Kaasalainen}, {Kryszczynska}, {Lamy}, {Le Coroller}, {Marciniak},
  {Michalowski}, {Michel}, {Santana-Ros}, {Tanga}, {Vigan}, {Witasse}, {Yang},
  {Antonini}, {Audejean}, {Aurard}, {Behrend}, {Benkhaldoun}, {Bosch},
  {Chapman}, {Dalmon}, {Fauvaud}, {Hamanowa}, {Hamanowa}, {His}, {Jones},
  {Kim}, {Kim}, {Krajewski}, {Labrevoir}, {Leroy}, {Livet}, {Molina},
  {Montaigut}, {Oey}, {Payre}, {Reddy}, {Sabin}, {Sanchez} and
  {Socha}}]{Vernazza2021}
\bibinfo{author}{{Vernazza}, P.}, \bibinfo{author}{{Ferrais}, M.},
  \bibinfo{author}{{Jorda}, L.}, \bibinfo{author}{{Hanu{\v{s}}}, J.},
  \bibinfo{author}{{Carry}, B.}, \bibinfo{author}{{Marsset}, M.},
  \bibinfo{author}{{Bro{\v{z}}}, M.}, \bibinfo{author}{{Fetick}, R.},
  \bibinfo{author}{{Viikinkoski}, M.}, \bibinfo{author}{{Marchis}, F.},
  \bibinfo{author}{{Vachier}, F.}, \bibinfo{author}{{Drouard}, A.},
  \bibinfo{author}{{Fusco}, T.}, \bibinfo{author}{{Birlan}, M.},
  \bibinfo{author}{{Podlewska-Gaca}, E.}, \bibinfo{author}{{Rambaux}, N.},
  \bibinfo{author}{{Neveu}, M.}, \bibinfo{author}{{Bartczak}, P.},
  \bibinfo{author}{{Dudzi{\'n}ski}, G.}, \bibinfo{author}{{Jehin}, E.},
  \bibinfo{author}{{Beck}, P.}, \bibinfo{author}{{Berthier}, J.},
  \bibinfo{author}{{Castillo-Rogez}, J.}, \bibinfo{author}{{Cipriani}, F.},
  \bibinfo{author}{{Colas}, F.}, \bibinfo{author}{{Dumas}, C.},
  \bibinfo{author}{{{\v{D}}urech}, J.}, \bibinfo{author}{{Grice}, J.},
  \bibinfo{author}{{Kaasalainen}, M.}, \bibinfo{author}{{Kryszczynska}, A.},
  \bibinfo{author}{{Lamy}, P.}, \bibinfo{author}{{Le Coroller}, H.},
  \bibinfo{author}{{Marciniak}, A.}, \bibinfo{author}{{Michalowski}, T.},
  \bibinfo{author}{{Michel}, P.}, \bibinfo{author}{{Santana-Ros}, T.},
  \bibinfo{author}{{Tanga}, P.}, \bibinfo{author}{{Vigan}, A.},
  \bibinfo{author}{{Witasse}, O.}, \bibinfo{author}{{Yang}, B.},
  \bibinfo{author}{{Antonini}, P.}, \bibinfo{author}{{Audejean}, M.},
  \bibinfo{author}{{Aurard}, P.}, \bibinfo{author}{{Behrend}, R.},
  \bibinfo{author}{{Benkhaldoun}, Z.}, \bibinfo{author}{{Bosch}, J.M.},
  \bibinfo{author}{{Chapman}, A.}, \bibinfo{author}{{Dalmon}, L.},
  \bibinfo{author}{{Fauvaud}, S.}, \bibinfo{author}{{Hamanowa}, H.},
  \bibinfo{author}{{Hamanowa}, H.}, \bibinfo{author}{{His}, J.},
  \bibinfo{author}{{Jones}, A.}, \bibinfo{author}{{Kim}, D.H.},
  \bibinfo{author}{{Kim}, M.J.}, \bibinfo{author}{{Krajewski}, J.},
  \bibinfo{author}{{Labrevoir}, O.}, \bibinfo{author}{{Leroy}, A.},
  \bibinfo{author}{{Livet}, F.}, \bibinfo{author}{{Molina}, D.},
  \bibinfo{author}{{Montaigut}, R.}, \bibinfo{author}{{Oey}, J.},
  \bibinfo{author}{{Payre}, N.}, \bibinfo{author}{{Reddy}, V.},
  \bibinfo{author}{{Sabin}, P.}, \bibinfo{author}{{Sanchez}, A.G.},
  \bibinfo{author}{{Socha}, L.}, \bibinfo{year}{2021}.
\newblock \bibinfo{title}{{VLT/SPHERE imaging survey of the largest main-belt
  asteroids: Final results and synthesis}}.
\newblock \bibinfo{journal}{Astronom. Astrophys.} \bibinfo{volume}{654},
  \bibinfo{pages}{A56}.
\newblock \DOIprefix\doi{10.1051/0004-6361/202141781}.
\bibitem[{{Vinatier} et~al.(2007a){Vinatier}, {B{\'e}zard}, {Fouchet},
  {Teanby}, {de Kok}, {Irwin}, {Conrath}, {Nixon}, {Romani}, {Flasar} and
  {Coustenis}}]{Vinatier2007a}
\bibinfo{author}{{Vinatier}, S.}, \bibinfo{author}{{B{\'e}zard}, B.},
  \bibinfo{author}{{Fouchet}, T.}, \bibinfo{author}{{Teanby}, N.A.},
  \bibinfo{author}{{de Kok}, R.}, \bibinfo{author}{{Irwin}, P.G.J.},
  \bibinfo{author}{{Conrath}, B.J.}, \bibinfo{author}{{Nixon}, C.A.},
  \bibinfo{author}{{Romani}, P.N.}, \bibinfo{author}{{Flasar}, F.M.},
  \bibinfo{author}{{Coustenis}, A.}, \bibinfo{year}{2007}a.
\newblock \bibinfo{title}{{Vertical abundance profiles of hydrocarbons in
  Titan's atmosphere at 15{\textdegree}S and 80{\textdegree}N retrieved from
  Cassini/CIRS spectra}}.
\newblock \bibinfo{journal}{Icarus} \bibinfo{volume}{188},
  \bibinfo{pages}{120--138}.
\newblock \DOIprefix\doi{10.1016/j.icarus.2006.10.031}.
\bibitem[{{Vinatier} et~al.(2007b){Vinatier}, {B{\'e}zard} and
  {Nixon}}]{Vinatier2007b}
\bibinfo{author}{{Vinatier}, S.}, \bibinfo{author}{{B{\'e}zard}, B.},
  \bibinfo{author}{{Nixon}, C.A.}, \bibinfo{year}{2007}b.
\newblock \bibinfo{title}{{The Titan $^{14}$N/ $^{15}$N and $^{12}$C/ $^{13}$C
  isotopic ratios in HCN from Cassini/CIRS}}.
\newblock \bibinfo{journal}{Icarus} \bibinfo{volume}{191},
  \bibinfo{pages}{712--721}.
\newblock \DOIprefix\doi{10.1016/j.icarus.2007.06.001}.
\bibitem[{{Vinatier} et~al.(2020){Vinatier}, {Math{\'e}}, {B{\'e}zard}, {Vatant
  d'Ollone}, {Lebonnois}, {Dauphin}, {Flasar}, {Achterberg}, {Seignovert},
  {Sylvestre}, {Teanby}, {Gorius}, {Mamoutkine}, {Guandique} and
  {Jennings}}]{Vinatier2020}
\bibinfo{author}{{Vinatier}, S.}, \bibinfo{author}{{Math{\'e}}, C.},
  \bibinfo{author}{{B{\'e}zard}, B.}, \bibinfo{author}{{Vatant d'Ollone}, J.},
  \bibinfo{author}{{Lebonnois}, S.}, \bibinfo{author}{{Dauphin}, C.},
  \bibinfo{author}{{Flasar}, F.M.}, \bibinfo{author}{{Achterberg}, R.K.},
  \bibinfo{author}{{Seignovert}, B.}, \bibinfo{author}{{Sylvestre}, M.},
  \bibinfo{author}{{Teanby}, N.A.}, \bibinfo{author}{{Gorius}, N.},
  \bibinfo{author}{{Mamoutkine}, A.}, \bibinfo{author}{{Guandique}, E.},
  \bibinfo{author}{{Jennings}, D.E.}, \bibinfo{year}{2020}.
\newblock \bibinfo{title}{{Temperature and chemical species distributions in
  the middle atmosphere observed during Titan's late northern spring to early
  summer}}.
\newblock \bibinfo{journal}{Astronom. Astrophys.} \bibinfo{volume}{641},
  \bibinfo{pages}{A116}.
\newblock \DOIprefix\doi{10.1051/0004-6361/202038411}.
\bibitem[{{Vinatier} et~al.(2012){Vinatier}, {Rannou}, {Anderson},
  {B{\'e}zard}, {de Kok} and {Samuelson}}]{Vinatier2012}
\bibinfo{author}{{Vinatier}, S.}, \bibinfo{author}{{Rannou}, P.},
  \bibinfo{author}{{Anderson}, C.M.}, \bibinfo{author}{{B{\'e}zard}, B.},
  \bibinfo{author}{{de Kok}, R.}, \bibinfo{author}{{Samuelson}, R.E.},
  \bibinfo{year}{2012}.
\newblock \bibinfo{title}{{Optical constants of Titan's stratospheric aerosols
  in the 70-1500 cm$^{-1}$ spectral range constrained by Cassini/CIRS
  observations}}.
\newblock \bibinfo{journal}{Icarus} \bibinfo{volume}{219},
  \bibinfo{pages}{5--12}.
\newblock \DOIprefix\doi{10.1016/j.icarus.2012.02.009}.
\bibitem[{{Vuitton} et~al.(2019){Vuitton}, {Yelle}, {Klippenstein}, {H{\"o}rst}
  and {Lavvas}}]{Vuitton2019}
\bibinfo{author}{{Vuitton}, V.}, \bibinfo{author}{{Yelle}, R.V.},
  \bibinfo{author}{{Klippenstein}, S.J.}, \bibinfo{author}{{H{\"o}rst}, S.M.},
  \bibinfo{author}{{Lavvas}, P.}, \bibinfo{year}{2019}.
\newblock \bibinfo{title}{{Simulating the density of organic species in the
  atmosphere of Titan with a coupled ion-neutral photochemical model}}.
\newblock \bibinfo{journal}{Icarus} \bibinfo{volume}{324},
  \bibinfo{pages}{120--197}.
\newblock \DOIprefix\doi{10.1016/j.icarus.2018.06.013}.

\end{thebibliography}

\bio{}
\endbio

\end{document}